\documentclass[%
      aps,
      prb, 
      twocolumn,
      superscriptaddress,
      showkeys]{revtex4-1}
      
\usepackage[utf8]{inputenc}
\usepackage[T1]{fontenc}
\usepackage{dcolumn}
\usepackage{bm}					
\usepackage{graphicx}
\usepackage{amsmath, amsfonts, amssymb, mathtools}	
\usepackage{latexsym} 				
\usepackage{xcolor}
\usepackage[caption = false]{subfig}
\usepackage[breaklinks, linkcolor = black]{hyperref}
\usepackage{tikz}
\usepackage{tabularx}
\usepackage{braket}
\usepackage{diagbox}
\usepackage[normalem]{ulem}
\usepackage{textcomp}
\usepackage{soul}
\usepackage{booktabs,eqparbox}
\graphicspath{{figures/}}

\begin{document}
\title{Reduced absorption due to defect-localized interlayer excitons in \\ transition metal dichalcogenide--graphene heterostructures}

%
 \author{Daniel \surname{Hernang\'{o}mez-P\'{e}rez}}

\email[]{daniel.hernangomez@weizmann.ac.il}

\affiliation{Department of Molecular Chemistry and Materials Science, Weizmann Institute of Science, Rehovot 7610001, Israel}

\author{Amir \surname{Kleiner}}

\affiliation{Department of Molecular Chemistry and Materials Science, Weizmann Institute of Science, Rehovot 7610001, Israel}

\author{Sivan \surname{Refaely-Abramson}}

\email[]{sivan.refaely-abramson@weizmann.ac.il}

\affiliation{Department of Molecular Chemistry and Materials Science, Weizmann Institute of Science, Rehovot 7610001, Israel}

\date{\today}

\begin{abstract}
Associating the presence of atomic vacancies to excited-state transport phenomena in two dimensional semiconductors is of emerging interest, and demands detailed understanding of the involved exciton transitions. Here we study the effect of such defects on the electronic and optical properties of WS\textsubscript{2}--graphene and MoS\textsubscript{2}--graphene van der Waals heterobilayers by employing many-body perturbation theory. We find that the combination of chalcogen defects and graphene adsorption onto the transition metal dichalcogenide layer can radically alter the optical properties of the heterobilayer, due to a combination of dielectric screening, the impact of the missing chalcogen atoms in the intralayer and interlayer optical transitions, and the different nature of each layer. By analyzing the intrinsic radiative rates of the most stable subgap excitonic features, we find that while the presence of defects introduces low-lying optical transitions, resulting in excitons with larger oscillator strength, it also decreases the optical response associated to the pristine-like transition-metal dichalcogenide intralayer excitons. Our findings relate excitonic features with interface design for defect engineering in photovoltaic and transport applications.
\end{abstract} 
 
 
\maketitle

Van der Waals heterostructures \cite{Wang2012, Geim2013, CastroNeto2016, Liu2016}, formed by vertically stacking  atomically-thin two-dimensional layers  through weak interlayer interaction, are considered one of the most promising systems for the next-generation of ultrathin optoelectronic and photovoltaic high-performance components with tunable properties and tailored functionalities modifiable at the atomic scale \cite{Radisavljevic2011, Andras2013, Ross2014, Hersam2014, Pospischil2014, Ferrari2016}.  
An important example of such heterostructures is the heterobilayer formed by stacking graphene \cite{Novoselov2005, Geim2009} with a monolayer transition metal dichalcogenide (TMDC) of the type XS\textsubscript{2}, where X is W, Mo \cite{Gierz2020, Gierz2021, Wang2021, Steinberg2021, Hernangomez2023, FariaJunior2023}. 
These are type I heterostructures which combine the high carrier mobility \cite{Mayorov2011}, high thermal conductivity \cite{Balandin2008} and semi-metallic character of graphene with the pseudospin circular dichroism \cite{Niu2008, Cao2012, Ye2017}, large quantum confinement, strong light absorption properties \cite{Bernardi2013} and sizeable spin-orbit interaction of a direct band gap TMDC \cite{Zhu2011, Lanzara2015}. 

The electronic and optical properties of layered TMDCs and their heterostructures are sensitive to the potential created by defects \cite{Lischner2018, RefaelyAbramson2018, RefaelyAbramson2019, barja2019identifying, Lischner2020}.
In particular, the most abundant and stable point defects in these systems are monoatomic chalcogen vacancies \cite{Idrobo2013, Robertson2013, Noh2014}. Electron-hole optical transitions between the defect and pristine states are known to produce novel sub-gap excitonic features \cite{Attaccalite2011, RefaelyAbramson2018, Mitterreiter2021, Hotger2021, Sigger2022, Micevic2022} and form localized excitons that were shown to intrinsically reduce the degree of valley polarization even without additional scattering mechanisms \cite{Koratkar2015, RefaelyAbramson2018, Mitterreiter2021, Amit2022, Hotger2023}.
Changes in the dielectric environment can impact the TMDC intrinsic light emission properties \cite{Steinhoff2015, Rana2016, Marie2016}. For instance, interlayer coupling between graphene and TMDC results in a notable quenching of excitonic photoluminescence \cite{Berciaud2020} or a widening of the exciton linewidth \cite{Heinz2017}. Engineering the exciton spontaneous decay time is also possible by microcavity formation by additional adsorbed layers and consequent Purcell effect \cite{Massicotte2016, Marie2019}. This effect has been shown to give low-temperature picosecond exciton photoresponses.
Therefore, and due to the defects ubiquitous nature, a  microscopic understanding of the emergent electronic and excitonic properties of TMDC--graphene (Gr) heterobilayers in the presence of vacancies is interesting for dynamical modelling of optoelectronic devices and applications.


In this work, we investigate the electronic and optical properties of WS\textsubscript{2}--Gr and MoS\textsubscript{2}--Gr heterobilayers with monoatomic chalcogen vacancies. We employ a GW-BSE approach \cite{Hedin1965, Hybertsen1985, Hybertsen1986, Albrecht1998, Rohlfing1998, Rohlfing2000, Deslippe2012} to compute the many-body electronic properties and optical characteristics and 
 find that due to the combination of screening and strong optical hybridization, absorption resonances of well-known pristine TMDC excitons are strongly quenched in the heterostructure, resulting in substantially altered absorbance properties compared to the pristine TMDC--Gr heterobilayer or defected TMDC without graphene. These pristine-like TMDC ``A'' and ``B'' peaks are largely reduced due to the mixing of the optical transitions with both graphene and defect electronic states and the electron-hole transitions determining their excitonic composition is fundamentally altered. This manifests also in a strong reduction of the excitonic binding energy for those absorption peaks. In addition, Fermi-level alignment of the defect transition levels and the magnitude of the spin-orbit interaction, determined by the choice of the TMDC, can result in additional substantial changes in the heterostructure optical properties. 
 We obtain the intrinsic radiative rates for excitons with the largest binding energy, which create strongly mixed subgap resonances, and show that the associated inverse rates are comparable to those calculated for pristine TMDC monolayers.  
 %

%
%
%
\begin{figure}
        \includegraphics[width=0.95\linewidth]{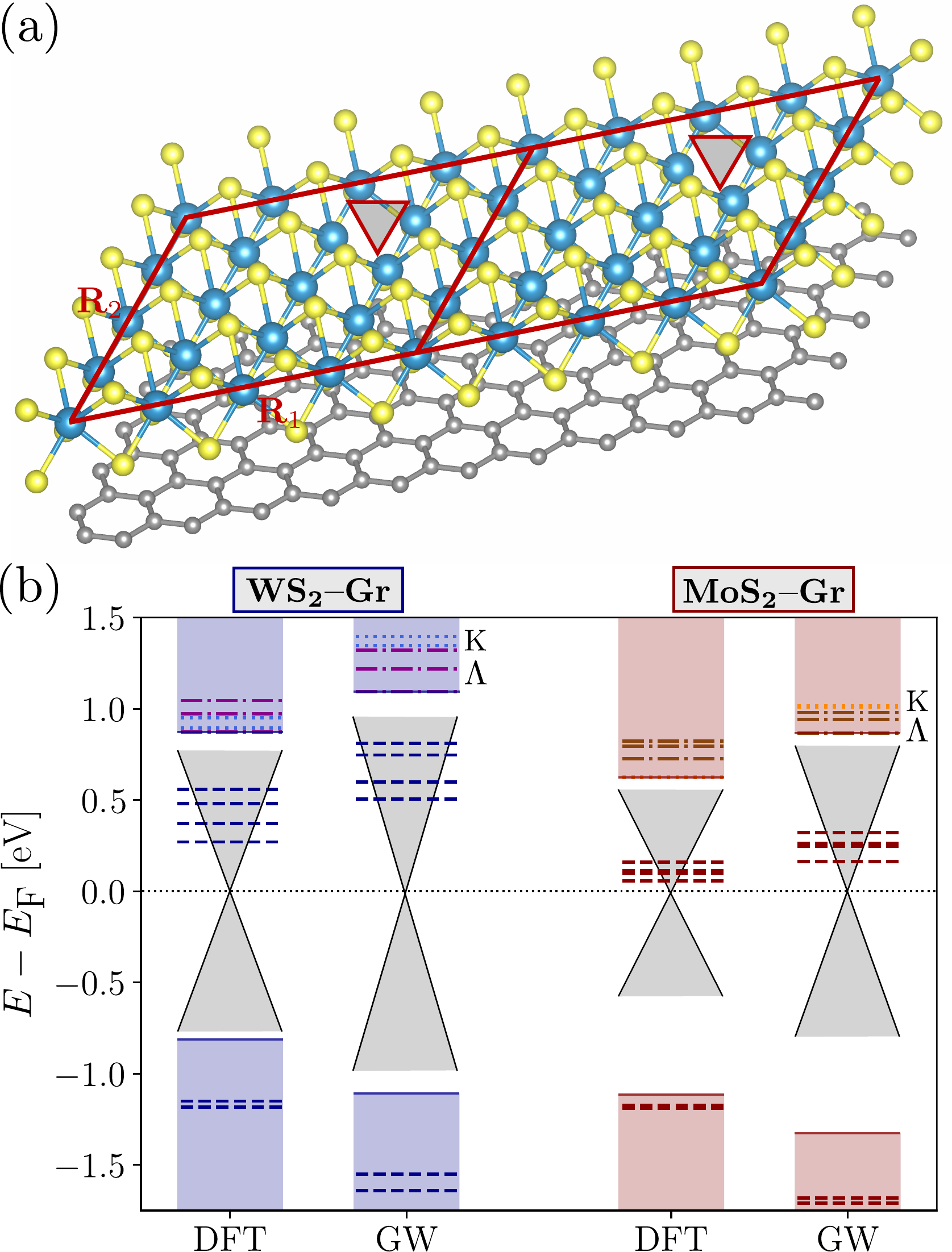}
    \caption{%
    (a) Schematics of a WS\textsubscript{2}--Gr heterobilayer. 
    Each supercell (two are shown here) forms a parallelepiped with lateral boundaries marked by the straight red lines (in-plane supercell lattice vectors are denoted by $\mathbf{R}_1$, $\mathbf{R}_2$). The monoatomic chalcogen vacancy position, located opposite to the graphene layer, is indicated by a red triangle.
    (b) DFT and GW calculated valence and conduction band energies at the $\bar{\textnormal{K}}$ point. The blue bars on the left-hand side correspond to WS\textsubscript{2}--Gr, the red bars on the right-hand side, to  MoS\textsubscript{2}--Gr. 
     The dashed lines denote the defect energy levels and the solid lines mark the top of the pristine TMDC valence and conduction bands. 
     Energies are related to the Fermi energy of the system defined as $E_\textnormal{F} = (E_{\textnormal{val}} + E_{\textnormal{cond}})/2$.
     The grey sketch of the Dirac cone represents the region of the TMDC pristine band gap occupied by graphene.
     Conduction levels with predominantly $\Lambda$ nature are represented by the dashed dotted lines, with those with predominantly $\textnormal{K}$ nature represented by dotted lines.
    }\label{f1} 
\end{figure}
We employ a commensurate supercell composed of $4 \times 4$ WS\textsubscript{2} (resp. MoS\textsubscript{2})  and $5 \times 5$ graphene elementary cells, see Fig. \ref{f1} (a) and Supporting Information (SI).  
We consider a vacancy concentration of $\sim 3$\% (at least one order of magnitude larger than the typical intrinsic vacancy concentration \cite{Abhay2019}) corresponding to a single monoatomic sulphur vacancy per supercell, located at the opposite side of the graphene layer. 
We first perform a geometry optimization of the supercell atomic positions, keeping the supercell volume constant (see \onlinecite{Hernangomez2023} and SI). This optimization reduces the nearest-neighbor bonds close to the vacancy (which shrinks and strains the TMDC lattice) as well as the interlayer distance between the TMDC and the graphene layer.
%
%
%
Using DFT as a starting point (with the PBE functional \cite{Perdew1996}), we perform a one-shot GW (G\textsubscript{0}W\textsubscript{0}) calculation for each TMDC--Gr heterobilayer (see computational details in the SI). 
Fig. \ref{f1} (b) shows the DFT and GW energies at the point $\bar{\textnormal{K}}$ of the supercell Brillouin zone in an energy level diagram. As expected based on previous studies \cite{Qiu2013, RefaelyAbramson2018, Jain2018}, GW increases the gap in both systems with the quasiparticle corrections, qualitatively conserving the DFT picture for these heterobilayers \cite{Hernangomez2023}.
There are four spin-orbit split defect states, shifted upon the GW calculation to higher energy with respect to the Fermi level. %
Far from the Fermi level, we find the valence band splitting due to spin-orbit interaction to be $\sim 460$ meV for WS\textsubscript{2}--Gr and $150$ meV for MoS\textsubscript{2}--Gr, which is consistent with $425 \pm 18$ meV and $170 \pm 2$ meV obtained from high-resolution ARPES measurements \cite{Lanzara2015}. 
The dielectric screening also shifts the pair of occupied defect states to lower energies (by $\sim 400-450$ meV for WS\textsubscript{2}--Gr and $\sim 400$ meV for MoS\textsubscript{2}--Gr).
Simultaneously, the pristine-like band gap of WS\textsubscript{2} increases from $1.69$ eV at the DFT level to $2.20$ eV at the GW level.
Similarly, the pristine-like band gap of MoS\textsubscript{2}--Gr changes from $1.73$ eV to $2.19$ eV.
These values reflect a significant renormalization of the TMDC band gaps compared to the monolayer case, as expected due to the quasi-metallic  character of graphene with GW corrections including image charge effects \cite{Jain2018}. 
Our GW results are in agreement with previous calculations with reported band gap reduction of $\sim 300-350$ meV \cite{Jain2018, Thygesen2015} compared to the pristine counterpart \cite{Qiu2013}. %
They are also well consistent with the experimental values of the quasi-particle bandgap found in MoS\textsubscript{2}--Gr heterostructures, reported to be $\sim 2.0-2.2$ eV \cite{Shi2015, Hersam2016}, and in WS\textsubscript{2}--Gr, which ranges $\sim 2.0-2.3$ eV \cite{Giusca2019, Gierz2021, Raja2017}. 

\begin{figure}
   \includegraphics[width=1.0\linewidth]{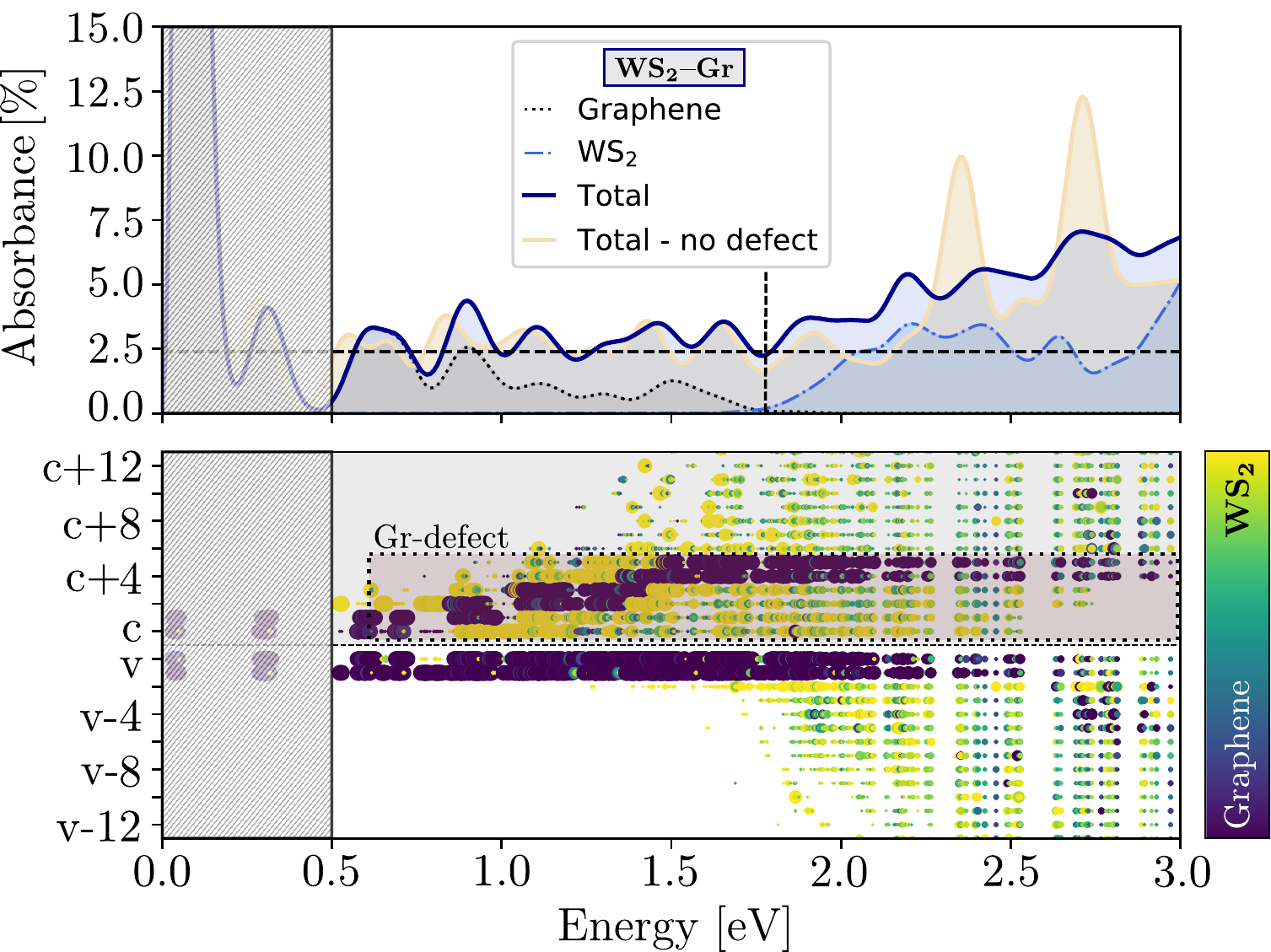}
    \caption{Absorbance and exciton contributions for the defected WS\textsubscript{2}--Gr heterobilayer. (Top) Absorbance calculated along one of the main in-plane polarization directions, as well as its decomposition into graphene and WS\textsubscript{2} contributions (interlayer contributions are read from the difference of the three traces).
    For comparison, we also show the absorbance of the pristine WS\textsubscript{2}-Gr heterobilayer (from \onlinecite{Kleiner2023}).
    The dashed horizontal black line marks the $2.4\%$ universal limit of graphene absorbance at infrared energies.
    The shaded box represents the estimated range for which we expect a smooth and monotonic spectrum dominated by graphene (instead of resonances resulting from finite \textbf{k}-grid sampling). The vertical dotted line marks the optical ranges below which excitons are dominated by defect-graphene sub-gap transitions to a range where excitons present larger intralayer TMDC composition.
    (Bottom) For each exciton composing the absorbance resonances, we represent the contribution of each electron and hole bands. Each dot corresponds to the band contribution to a given exciton summed over all $\mathbf{k}$ points %
    (only bright contributions whose 
    oscillator strength are $> 5$ a.u. are shown). For clarity, all %
    dots with value $\geq 10^3$ a.u. have the same area. 
    The color code corresponds to the layer composition  of each contribution and the dotted box marks the graphene-defect empty bands. 
    }\label{f5} 
\end{figure}

%
%
 
%
%
Next, we examine the excitonic properties of defected  WS\textsubscript{2}--Gr and MoS\textsubscript{2}--Gr heterobilayers using the Bethe-Salpether equation \cite{Rohlfing1998, Rohlfing2000} within the Tamm-Dancoff approximation (see SI).
We show in Fig. \ref{f5}, top panel, the absorbance of WS\textsubscript{2}--Gr as well as its decomposition into intralayer graphene, intralayer TMDC and interlayer contributions (see SI for the case of MoS\textsubscript{2}--Gr). 
At low optical energies, intralayer graphene electron-hole excitonic features are found to be most prominent. Graphene intraband transitions (not considered here, as well as temperature effects) are known to dominate this regime \cite{Li2009}, marked by a dashed grey rectangle. The resonant peaks are a consequence of the finite \textbf{k}-grid sampling of the graphene Dirac cone and therefore, the absorption in this region is actually continuous in the dense grid limit.
In the high infrared spectral range, excitonic peaks corresponding to interlayer graphene--defect and graphene--pristine optical transitions appear  at higher energies while graphene intralayer contributions become less relevant. At optical energies $\gtrsim 2.0$ eV, intralayer TMDC contributions (in the form of defect-defect, defect-pristine and pristine-pristine band transitions) become the dominant features of the spectrum over the interlayer contributions.
%
Out of all optical transitions sampled for our \textbf{k}-grid and energy range, 0.4\% belong to intralayer graphene while 62\% correspond to intralayer TMDC transitions, with the remaining 37\% representing a large degree of interlayer mixing. 

The spectral absorbance of graphene for infrared light is almost constant at $2.4\%$ \cite{Li2009, Nair2008, Heinz2008, Gusynin2006, Peres2006, Carbotte2008, Stauber2008}. This limit is represented by a dashed horizontal line in Fig. \ref{f5}. We find that absorbance resonances in the optical range between $\sim 1.0-1.6$ eV oscillate around values larger than the graphene infrared constant limit. These features are likely due to excitons that involve defects and should persist in the dense grid limit.  
The computed absorbance values in this range are also consistent with those calculated for defected MoSe\textsubscript{2}  in the absence of graphene \cite{RefaelyAbramson2018}.
To further validate our findings, we compare the absorbance for WS\textsubscript{2}--Gr with and without vacancies. We observe that, unlike the defected heterostructure, the absorbance of the pristine heterobilayer oscillates around the graphene  limit in the same energy range , supporting our previous conclusion.
In the visible range (\textit{i.e.} $\sim 1.6-3.2$ eV), experiments on WS\textsubscript{2} have reported a red shift and a significant quenching of the excitonic resonances in photoluminescence upon graphene adsorption \cite{Heinz2017, Magnozzi2020}. These effects were attributed to changes in the dielectric environment  of the TMDC due to the adsorbed quasi-metallic layer.
In defected WS\textsubscript{2}--Gr heterobilayers, we observe 
a reduction of the strength of the pristine-like TMDC resonances in addition to the strong interlayer mixing in the sub-gap optical region associated with the defect states. This effect is attributed to the strong optical mixing of TMDC and graphene, which results in additional interlayer optical transitions and redistribution of the oscillator strength due to the defects.
These combined effects of the graphene and vacancies quench the pristine-like ``A'' and ``B'' resonances \cite{Ramasubramaniam2012, Druppel2018} and also broaden them \cite{Amit2022}. As a consequence, they are no longer as dominant in the spectrum. Moreover, the composition of the absorption peaks, clearly defined at $\sim 2.2$, $\sim 2.4$ and $\sim 2.7$ eV, also changes drastically compared to the expectation for the pristine or defected TMDC monolayers (see SI).

 \begin{figure*}
   \includegraphics[width=1.0\linewidth]{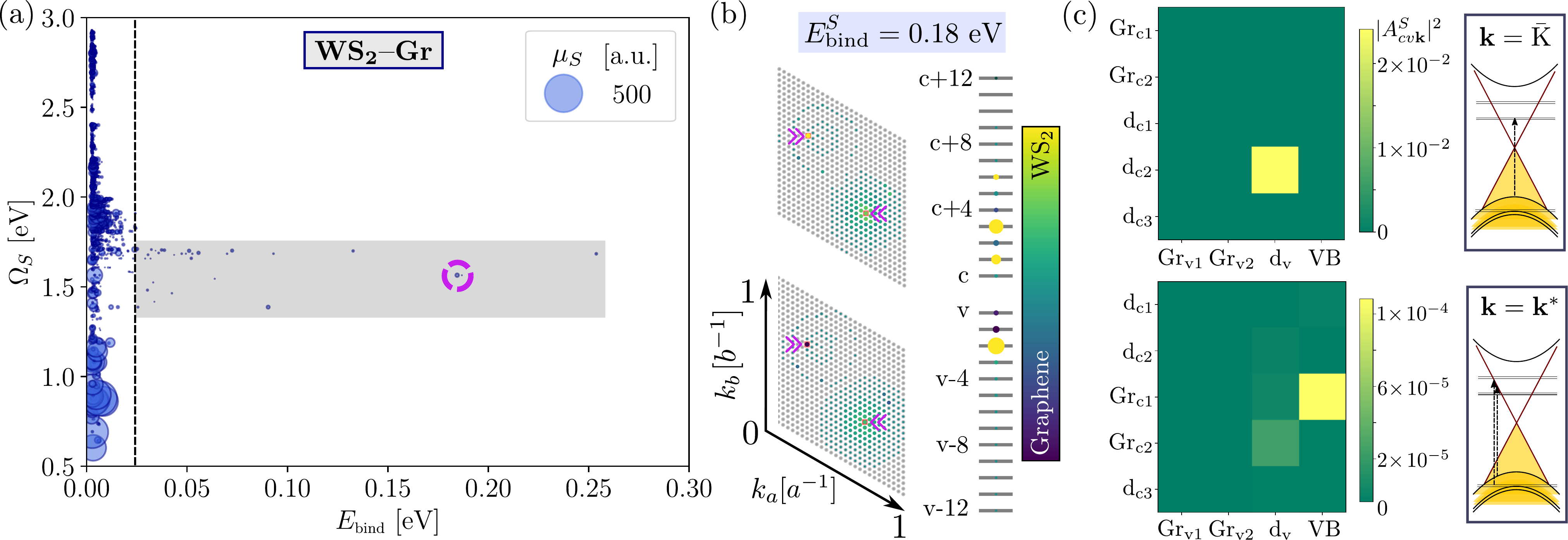}
    \caption{(a) Exciton energies for WS\textsubscript{2}--Gr, $\Omega_S$, as a function of their binding energy, $E_\textnormal{bind}$.
    Only excitons for which $E_\textnormal{bind} > 2.5$ meV are shown ($\sim 8000$ out of $142 884$ excitons for our \textbf{k}-grid sampling and bands). The size of each dot is proportional  to the oscillator strength, $\mu_S$. %
    Bright excitons, in particular those dominated by intralayer graphene transitions have very small binding energies (smaller than the thermal energy at room temperature, marked by a dashed black vertical line).
    Excitons within the energy range where the pristine ``A'' and ``B'' features would be expected are dark and also have a very small binding energy.    
    The grey rectangle corresponds to  excitons with  binding energies  larger than $25$ meV. 
    (b) Brillouin zone exciton distribution and transition band diagram for the exciton marked with a purple circle in  (a). 
    The top Brillouin zone represents transitions to the conduction bands; the bottom one, transitions from the valence bands. 
    (c) Transition band diagram and sketch of the optical transitions at selected \textbf{k}-points  marked with purple arrows in (b). 
    }\label{f8} 
\end{figure*}
%

To further understand this effect, we show in Fig. \ref{f5}, bottom panel, the contribution of each band to the exciton (similarly to our previous analysis of exciton state mixing in defected systems \cite{RefaelyAbramson2018, Mitterreiter2021, Steinitz2022}).
Each excitonic state, $|\Psi^S \rangle$, defined by its wavefunction amplitude $A^S_{vc\mathbf{k}}$ and energy $\Omega_S$, is represented by a column of dots whose area is proportional to $\sum_{v\mathbf{k}}|A^S_{vc\mathbf{k}}|^2$, for each electron ($c$), and $\sum_{c\mathbf{k}}|A^S_{vc\mathbf{k}}|^2$, for each hole ($v$).
The color of the dot represents the target layer from which ($c$) or to which ($v$) the transitions occur. 
We observe intralayer graphene optical transitions in the low energy region ($\lesssim 0.5$ eV), while excitonic resonances with interlayer character appear only above $0.5$ eV.
The dispersive nature of graphene can be seen from the increase of the conduction band number for graphene with the energy.
As expected, the quenched high-energy resonances %
show significant contribution of mixed TMDC and graphene optical transitions, thus, although they appear in similar positions they no longer possess true ``A'' and ``B'' characters (see SI). We note that while graphene reduces absorption properties without the defects as well \cite{Kleiner2023}, here the effect is further pronounced due the electron-hole defect and non-defect mixing in the sub-gap region.  
%
%
%
%
%
%
%

%
The binding energy of the exciton quantifies how strongly bound are the electrons and holes participating in the excitation. 
Intralayer graphene excitonic features have vanishing small binding energies ($\sim 0-1$ meV, see also \onlinecite{Li2009, Li2011}) despite their strong oscillator strength. This differs significantly from pristine or encapsulated TMDC excitons, which have large oscillator strength and binding energies.
In particular, experimental estimations of the binding energies are $0.3-0.4$ eV for MoS\textsubscript{2}, and $0.3-0.7$ eV for WS\textsubscript{2} pristine ``A'' excitons \cite{Urbaszek2018}. %
Theoretical predictions give $0.6$ eV for MoS\textsubscript{2} and $0.2$ eV for WS\textsubscript{2} pristine ``B'' excitons \cite{Druppel2018}. %
In Fig. \ref{f8} (a), we present the exciton energy as a function of binding energy for WS\textsubscript{2}--Gr (see SI for MoS\textsubscript{2}--Gr). For excitons in the optical region where the pristine ``A'' and ``B''  resonances  would be expected, we observe that excitons have substantially lower binding energies compared to the pristine or encapsulated counterparts, as well as substantially smaller oscillator strength. We attribute this to a redistribution of the oscillator strength due to the combined effect of substantial hybridization of the defect and non-defect electron-hole transitions with graphene, as well as the small binding properties of excitons in graphene due to its quasi-metallic nature at low energies.
Importantly, we also find excitons (with oscillator strength in the range $10^{-2}-1.0$ a.u.) in the optical region $1.5-2.0$ with a binding energy comparable to that of pristine excitons in the absence of graphene layer. These excitons result from intralayer optical transitions to defect states and interlayer graphene-defect transitions.
To gain insight into the nature of these excitons, we show in Fig. \ref{f8} (b) the \textbf{k}-space distribution for a representative  case  marked by a circle in panel (a). 
It is worth noting that their degree of localization cannot be used to infer  the strength of the binding, as excitons with similar
binding  may exhibit optical transitions in very different regions of the Brillouin zone due to the dispersive nature of the graphene bands and the delocalization of defect states in \textbf{k}-space.
Fig. \ref{f8} (c) displays the optical transitions at selected points in the Brillouin zone, supporting our analysis that this exciton is formed by a combination of defect-defect (notably at $\bar{\textnormal{K}}$), graphene-valence and graphene-defect transitions (for the \textbf{k}-point noted as $\mathbf{k}^\ast$).

Finally, we relate our findings to the intrinsic radiative decay rates of zero-momentum excitons, which can be computed from the GW-BSE oscillator strength and excitation energy \cite{Spataru2005, Palummo2015, Bernardi2018}. We consider the inverse rate, which scales as $\gamma_S^{-1}:=\tau_S  \sim \Omega_S/\mu_S$.
This rate accounts only for part of the radiative linewidth, as other contributions, \textit{e.g.} electron-phonon and exciton-phonon terms are not included, and is useful to evaluate the significance of the oscillator strength. 
Our analysis reveals that the inverse rates for the excitons with binding $>50$ meV can be as large as $\tau_S \sim 0.1$ ps for both heterobilayers. However, depending on the oscillator strength, they can be shorter, even as small as $\tau_S \sim 0.1$ fs for MoS\textsubscript{2}--Gr (see SI). Intralayer graphene excitonic features, which have significantly small binding, have substantially larger intrinsic rates due to their large optical transition dipole. 
Dark interlayer excitons with large binding have larger inverse rates, as large as $\sim 100$ ps for WS\textsubscript{2}--Gr, due to the smaller oscillator strength.
For WS\textsubscript{2}--Gr without defects bright ``A'' and ``B'' excitons  in the TMDC layer, $\tau_S$ can be even shorter, essentially due to the increased oscillator strength (between two and three orders of magnitude compared to the defect-related excitons, see SI) which yields $\tau_S \sim 10^{-4}-10^{-5}$ fs. We note that compared to pristine TMDCs \cite{Palummo2015}, graphene adsorption has a strong impact on $\tau_S$, which only become comparable again to the pristine ones in the presence of defects due to the strong exciton hybridization of the graphene and the subgap vacancy-related features.
Furthermore, charge transfer times of photocarriers at TMDC--graphene interfaces \cite{Jin2018, Yuan2018, Gierz2020, Gierz2021, Wang2021}, where single-particle defect tunneling is understood to be the dominating coherent transport channel \cite{Gierz2021, Hernangomez2023} can be of similar order of magnitude. In this scenario, defects slow down coherent charge transfer due to relatively small interlayer tunneling. Interestingly, in the presence of graphene, defects optically enhance transitions associated to them, resulting in  excitons with significantly higher oscillator strength, compared to the reduced oscillator strength of the original pristine-like ``A'' and ``B'' TMDC excitons.

%
%

%
In conclusion, we have studied the electronic and optical properties of WS\textsubscript{2}--Gr and MoS\textsubscript{2}--Gr heterobilayers with chalcogen vacancies employing first-principles many-body perturbation theory. 
We find that strong hybridization of the defect states with graphene gives rise to subgap features, which manifest as strong resonances in the optical absorbance spectrum, while quenching the  ``A'' and ``B'' pristine exciton peaks originally coming from intralayer TMDC transitions. These altered absorption features may be used to extend the functionality in the infrared of solar cells. We have analyzed the stability of the excitons and found a strong reduction of the binding energy for those TMDC excitons, while strongly hybridized interlayer and defect-dominated excitons have binding energies up to $\sim 250$ meV.
We computed the intrinsic radiative decay rate of these excitons and found inverse rates of up to $0.1$ ps. Overall, our results demonstrate how point-like defects can be used to design optical features in graphene-based van der Waals heterostructures, where excitons inherit properties from two well-distinct layers in a non-trivial way, pointing to the relevance of a first-principles understanding of many-body effects in the description of these systems for transport and potential optoelectronic applications.

%
\subparagraph{Acknowledgments.} 
The authors acknowledge Tomer Amit, Mar\'ia Camarasa--G\'omez, Alexey Chernikov, Florian Dirnberger, Paulo E. Faria Junior and Alexander Holleitner for insightful discussions.
The authors are thankful to Simone Latini, Lede Xian and \'Angel Rubio for the initial geometry employed as a starting point of the calculations performed in this manuscript.
The computations were carried out in the Chemfarm local cluster at the Weizmann Institute of Science and the Max Planck Computing and Data Facility cluster. 
D. H.-P. and A. K. acknowledge a Minerva Foundation grant 7135421. This research is supported by the German Research Foundation (DFG) through the Collaborative Research Center SFB 1277 (Project-ID 314695032, project B10). S. R. A. is an incumbent of the Leah Omenn Career Development Chair.
This project has received funding from the European Research Council (ERC), Grant agreement No. 101041159.

%
\subparagraph{Supporting Information.}
The Supporting Information is available free of charge on the ACS Publications website at \texttt{http://pubs.acs.org}.\\
Computational details and methods, additional results, convergence checks, absorbance and additional results for MoS\textsubscript{2}--Gr, wavefunction density of relevant conduction states, exciton reciprocal space composition for relevant energies in the WS\textsubscript{2}--Gr spectra, additional figures.


\bibliography{biblio}

\end{document}


\title{{Supporting Information \\ Reduced absorption due to defect-localized interlayer excitons in \\ transition metal dichalcogenide--graphene heterostructures}}

\author{Daniel \surname{Hernang\'{o}mez-P\'{e}rez}}


\affiliation{Department of Molecular Chemistry and Materials Science, Weizmann Institute of Science, Rehovot 7610001, Israel}

\author{Amir \surname{Kleiner}}

\affiliation{Department of Molecular Chemistry and Materials Science, Weizmann Institute of Science, Rehovot 7610001, Israel}

\author{Sivan Refaely-Abramson}


\affiliation{Department of Molecular Chemistry and Materials Science, Weizmann Institute of Science, Rehovot 7610001, Israel}

\keywords{2D materials, transition-metal dichalcogenides, heterostructures, defects, graphene, excitons}

\maketitle

\begin{center}
 \vspace{-.6cm} 
 Email: daniel.hernangomez@weizmann.ac.il \\ \hspace{1.8cm} sivan.refaely-abramson@weizmann.ac.il
 \end{center}

\tableofcontents
\clearpage

\section{Computational details and methods}\label{app:computational}

\begin{figure}[h]
   \includegraphics[width=0.6\linewidth]{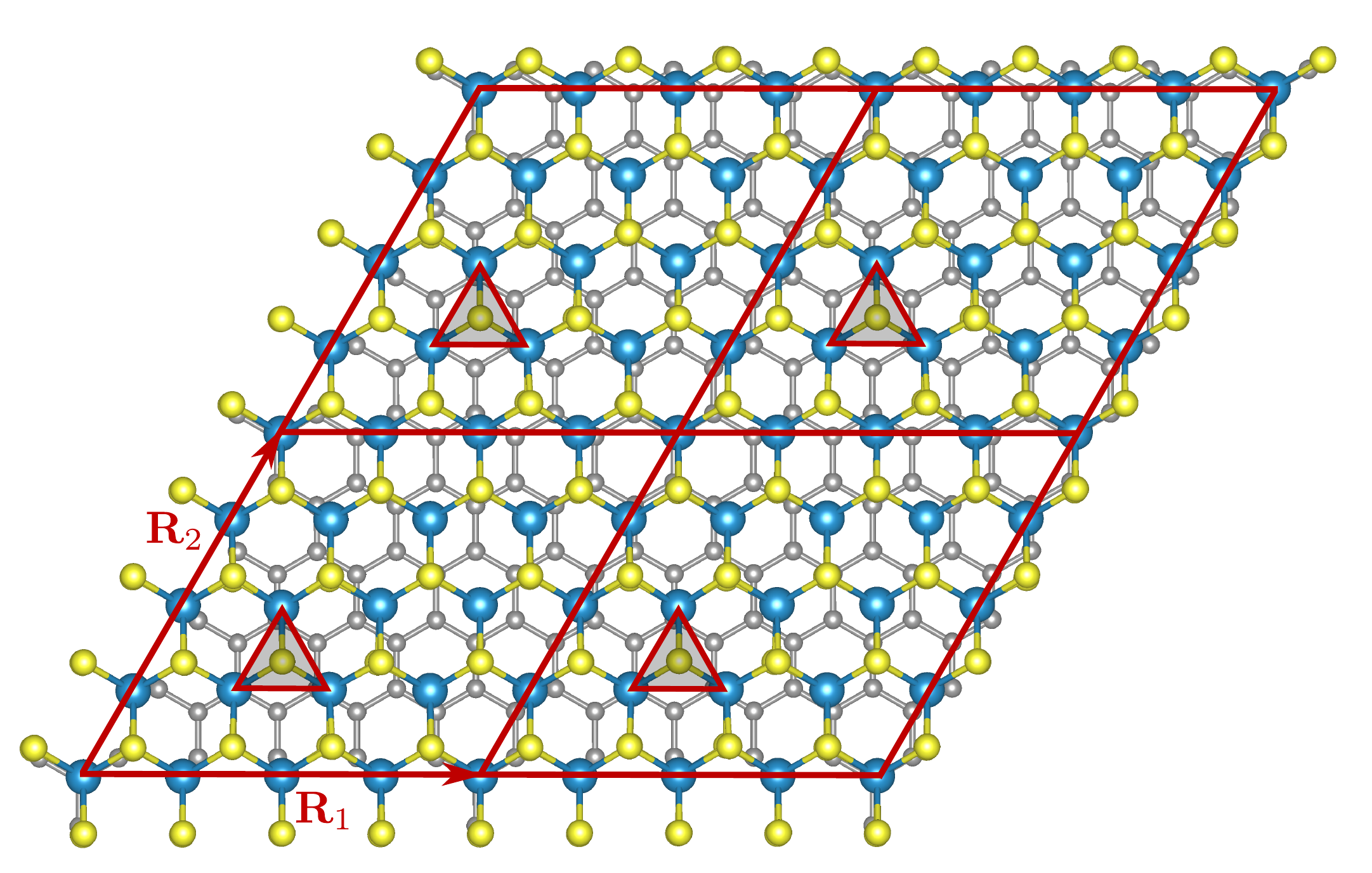}

    \caption{Top view of the WS\textsubscript{2}--Gr supercell. Four supercells are shown, each supercell forms a parallelepiped whose lateral boundaries are given by the straight red lines (the in-plane supercell lattice vectors are labeled as $\mathbf{R}_1$, $\mathbf{R}_2$). The chalcogen vacancy, located in the TMDC layer on the opposite side of the graphene layer, is indicated by a red triangle.
    }\label{f5} 
\end{figure}

\subparagraph{Geometry.} As a starting point for the supercell optimization in the presence of the vacancy (see details below), we consider the geometry of a pristine heterobilayer  with a lattice parameter whose length corresponds to {an average} nearest-neighbor metal-metal distance (equal to the in-plane monolayer TMDC lattice constant) of $\bar{d}_{\textnormal{X--X}} = 3.15 $\,\AA \, both for MoS\textsubscript{2} and WS\textsubscript{2}. 
This value is almost equal to the experimental monolayer lattice parameter of both TMDC (MoS\textsubscript{2}; 3.15 \AA\, \onlinecite{Nicklow1975} and WS\textsubscript{2}; 3.153 \AA\, \onlinecite{Schutte1987}).
As stated in the main text, the supercell is composed of $4\times4$ TMDC unit cells and $5\times 5$ graphene unit cells. Therefore, it is made of $97$ atoms and possesses a rhomboedral shape with in-plane lattice vectors of length $|\mathbf{R}_{1,2}| \simeq 12.6$ \AA\, (see Fig. \ref{f5}).
The vacuum distance between the periodic replicas in the out-of-plane direction was taken to be $\sim 10$ \AA\, and the average interlayer distance within the supercell was $\bar{d}_\textnormal{inter} = 3.43$ \AA\, {both for Mo and W}.

\subparagraph{Density functional theory.} The DFT calculations were performed employing the implementation of DFT of \textsc{Quantum Espresso}. \cite{Giannozzi2009, Giannozzi2017} %
We used the non-empirical PBE generalized gradient approximation for the exchange--correlation functional \cite{Perdew1996}.
We employed a plane-wave basis set and included spin-orbit interaction by means of full relativistic norm-conserving pseudopotentials \cite{Dojo2019}. We considered a basis cut-off of $50$ Ry for both TMDC--Gr interfaces.
The self-consistent charge density was converged within a $6 \times 6 \times 1$ \textbf{k}-grid, with Fermi-Dirac smearing of $10^{-5}$ Ry for fractional occupations. The calculation was considered to be converged only if the total energy difference between consecutive iterations within the self-consistent field cycle was smaller than the threshold value of $10^{-9}$ Ry.

The supercells were initially preoptimized with VASP \cite{Kresse1996} in the absence of chalcogen vacancies. For the exchange-correlation functional a local density approximation (LDA) was employed, with a basis set energy cut-off of $600$ eV. The self-consistent charge density for the geometry relaxations was converged in a $6\times 6 \times 1$ \textbf{k}-grid as well. 
The supercells were subsequently relaxed, fixing only the position of the supercell lattice vectors and optimizing the position of the TMDC atoms within the supercell after the chalcogen atom was removed. 
The position of the atoms was relaxed until all components of the forces were smaller than a threshold value of $10^{-3}$ Ry/$a$\textsubscript{0}.
This second optimization was done with \textsc{Quantum Espresso}, using PBE and the van-der Waals corrected functional \texttt{vdw-df-09} \cite{Thonhauser2007, Troy2009, Berland2015} to properly account for changes in the interlayer separation.

\subparagraph{GW.} Using the DFT wavefunctions and energies as a starting point, we computed the corrected quasi-particle energy spectrum by performing a one-shot non-self-consistent GW calculation (G\textsubscript{0}W\textsubscript{0}). Our GW calculations were performed with the package Berkeley{GW}, including spin-orbit interaction \cite{Deslippe2012, Rohlfing2000, Wu2020, Louie2022}.
The dielectric function was obtained with the generalized plasmon-pole model of Hybertsen-Louie \cite{Hybertsen1986}.
We employed a cut-off of $5$ Ry in the dielectric screening and a total of \textcolor{black}{$2499$} states for the summation over the occupied and unoccupied states.
We used a non-uniform neck subsampling scheme to sample the Brillouin zone close to $|\mathbf{q}| \rightarrow 0$ and speed up  the convergence with respect to the 
\textbf{k}-grid sampling \cite{Qiu2017}. Within this scheme, we considered a $6\times 6 \times 1$ uniform \textbf{q}-grid and $10$ additional \textbf{q}-points around $\mathbf{q}= \mathbf{0}$.
A truncated Coulomb interaction was considered in the perpendicular direction to the heterostructure to prevent spurious interactions between the periodic replicas in this direction \cite{Ismail2006}. \textcolor{black}{This set of parameters yields converged quasiparticle gaps within 100 meV.}

\subparagraph{BSE.}
To study the excitonic features, we solved the Bethe-Salpether equation (BSE) \cite{Rohlfing1998, Rohlfing2000} 
\begin{equation}
    (E_{c\mathbf{k}} - E_{v\mathbf{k}}) A_{vc\mathbf{k}}^S + \sum_{v'c' \mathbf{k}'} K^{\textnormal{eh}}_{vc\mathbf{k}; v'c'\mathbf{k}'} A_{v'c'\mathbf{k}'}^S  = \Omega_S A_{vc\mathbf{k}}^S,\label{e3}
\end{equation}
where $E_{c\mathbf{k}}$ (resp. $E_{v\mathbf{k}}$) are the quasiparticle energies of the conduction (resp. valence) bands, $K^{\textnormal{eh}}_{vc\mathbf{k}; v'c'\mathbf{k}'}  = \langle vc \mathbf{k}|\hat{K}^{\textnormal{eh}}|v'c' \mathbf{k}' \rangle$ are the matrix elements of the electron-hole interaction kernel, defined from the addition of an attractive screened direct and a repulsive bare exchange Coulomb contributions, $\Omega_S$ is the exciton energy and $A^S_{vc\mathbf{k}}$ is the amplitude of the exciton state $|\Psi^S \rangle$.
This equation sets an eigenvalue problem, $\hat{H}^\textnormal{BSE}|\Psi^S \rangle = \Omega_S |\Psi^S \rangle$, where the matrix elements of the BSE Hamiltonian in the electron-hole basis are given by
\begin{equation}
    {H}^\textnormal{BSE}_{cv\mathbf{k};c'v'\mathbf{k}'} = (E_{c\mathbf{k}} - E_{v\mathbf{k}}) \delta_{c,c'}\delta_{v,v'}\delta_{\mathbf{k},\mathbf{k}'} + K^{\textnormal{eh}}_{vc\mathbf{k}; v'c'\mathbf{k}'}.
\end{equation}
This representation of the BSE assumes that the (real-space) direct exciton wavefunction is described as the coherent superposition of electrons and holes at each $\mathbf{k}$-point,
\begin{equation}
   \langle \mathbf{r}_e, \mathbf{r}_h | \Psi^S \rangle = \sum_{vc\mathbf{k}} A^S_{vc\mathbf{k}}  \psi^\ast_{v\mathbf{k}}(\mathbf{r}_h) \psi_{c\mathbf{k}}(\mathbf{r}_e), 
\end{equation}
with $\psi_{c\mathbf{k}}(\mathbf{r}_e)$ being the spinor wavefunction describing the electron at position $\mathbf{r}_e$ with conduction band quantum number $c$ and crystal momentum $\mathbf{k}$ (correspondingly, $\psi_{v\mathbf{k}}(\mathbf{r}_h)$ is the spinor wavefunction describing a hole at position $\mathbf{r}_h$ and characterized by the valence band quantum number $v$ and same crystal momentum $\mathbf{k}$).

Eq. \eqref{e3} was solved using the Berkeley{GW} package \cite{Deslippe2012, Rohlfing2000, Wu2020}. 
The matrix elements were computed on a Monkhorst-Pack  $9 \times 9 \times 1$ \textbf{k}-grid and the result interpolated to a uniform \textcolor{black}{$27 \times 27 \times 1$ \textbf{k}-grid} that we employ in any subsequent analysis and calculation.
We employed the Tamm-Dancoff approximation and evaluated the Coulomb interaction kernel for all possible transitions between pairs of bands $(n,m) \rightarrow (n', m')$, with an energy cut-off of $5$ Ry for the dielectric matrix within the electron-hole kernel matrix elements.
We considered for the main paper a total of \textcolor{black}{$28$ bands ($14$ valence and $14$ conduction bands)} in the absorption calculations, which include both the defect bands as well as all the relevant low energy pristine valence and conduction bands of the heterobilayer.
\textcolor{black}{These parameters converged the calculated excitonic spectra within 100 meV for the defect-dominated subgap features and below 10 meV for the most prominent absorption resonances in the region where intralayer non-defect TMDC excitons become more relevant.}
In the absorption calculation, we also avoid the heavy calculation of the $\mathbf{q}$-shifted wavefunctions on the interpolation grid by evaluation of the matrix elements of the velocity operator, $\hat{\mathbf{v}}$, instead of the momentum, $\mathbf{p} = \mathfrak{i} \nabla$. This involves neglecting terms in the sums proportional to $|\langle 0 |[\hat{V}_{\textnormal{ps}}, \hat{\mathbf{r}}]| S \rangle|^2$, where $\hat{V}_{\textnormal{ps}}$ is the non-local part of the pseudopotential \cite{Rohlfing2000, Deslippe2012}. Including the non-local terms has been shown to not qualitatively change the shape of water X-ray absorption spectrum \cite{Qiu2022}.

\subparagraph{Projected density of states.} We compute the layer contribution of each band to a given exciton state, first by obtaining the \textbf{k}-projected density of states (DoS) of each layer, $l = \{\textnormal{WS}_2, \textnormal{Gr}\}$ from the \textbf{k}-resolved projected DoS
\begin{equation}\label{e1}
    g^{l}_{n\mathbf{k}}(E) = \sum_{\{i_A,A\} \in l} |\langle \phi_{i_A}^A| \psi_{n\mathbf{k}} \rangle|^2 \delta(E - E_{n\mathbf{k}}),
\end{equation}
where $|\psi_{n\mathbf{k}} \rangle $ and $E_{n\mathbf{k}}$ are the Kohn-Sham states and energies and the sum runs over atoms $A$ and orbitals $i_A$ of the corresponding layer $l$.
We further normalize this quantity for each layer as $\tilde{g}(E) = g(E) / \textnormal{max}[g(E)]$ so that 
\begin{equation}\label{e2}
\tilde{g}_{n\mathbf{k}}(E) = \tilde{g}^{\textnormal{WS}_2}_{n\mathbf{k}}(E) - \tilde{g}^{\textnormal{Gr}}_{n\mathbf{k}}(E),
\end{equation}
 is defined in the range $[-1,1]$. This way, $\tilde{g}_{n\mathbf{k}}(E) = -1$ corresponds exclusively to graphene contribution and $\tilde{g}_{n\mathbf{k}}(E) = 1$  exclusively to TMDC contribution.

 \subparagraph{Heterostructure decomposition.} 
 We employ Eq. \eqref{e2} to display the color of the band contributions to the exciton decomposition  as well as the absorbance decomposition into intralayer and interlayer parts in the main paper. In particular, for Fig. 2 and Fig. \textbf{S5}, the contributions to a given conduction band are obtained as $\sum_{vc\mathbf{k}} \tilde{g}_{v\mathbf{k}}|A^S_{vc\mathbf{k}}|^2$ while the contributions to a given valence band result from $\sum_{vc\mathbf{k}} \tilde{g}_{c\mathbf{k}}|A^S_{vc\mathbf{k}}|^2$. 
 The \textbf{k}-resolved decomposition in Fig. 3(b) of the main paper (see also Figs. \ref{f18}, \ref{f7}, \ref{f8}) are obtained using the same expressions but without the summation over the crystal momentum.

\subparagraph{Absorbance.} 

From the absorption, we compute the associated absorbance using\cite{Li2009}
\begin{equation}
A(\omega) = \dfrac{\omega L_z}{c}  \epsilon_2(\omega),
\end{equation}
where $\omega$ is the photon frequency, $c$  the speed of light, $L_z$  the distance between the heterostructure and its periodic replicas and $\epsilon_2(\omega)$  the imaginary part of the dielectric function.

\subparagraph{Exciton binding energy.} 
The excitonic binding energies are calculated from the difference between the expectation values of the diagonal and the full BSE Hamiltonian,
\begin{align}
    E^S_\textnormal{bind} &=
    \langle \Psi^S | \hat{H}^{\textnormal{BSE}} - \hat{K}^{\textnormal{eh}} | \Psi^S \rangle - \langle \Psi^S | \hat{H}^{\textnormal{BSE}} | \Psi^S \rangle, \notag \\
    &=\sum_{vc\mathbf{k}}|A^S_{vc\mathbf{k}}|^2 (E_{c\mathbf{k}} - E_{v\mathbf{k}}) - \Omega_S,\label{eq:binding}
\end{align}
where $\hat{H}^{\textnormal{BSE}}$ is the BSE Hamiltonian and  $\hat{K}^{\textnormal{eh}}$ the electron-hole interaction kernel; $E_{\alpha\mathbf{k}}$, the quasi-particle bands; and $\Omega_S$, the exciton energy.

\subparagraph{Intrinsic decay rate.}
To compute the intrinsic decay rate of the zero-momentum excitons we follow Refs. \onlinecite{Spataru2005, Palummo2015, Bernardi2018}
 \begin{equation}
      \gamma_S = \dfrac{4 \pi e^2}{m^2 c} \dfrac{\mu_S}{A_{c} \Omega_S},
 \end{equation}
 where $m$ is the electron mass, $\mu_S$ is the BSE oscillator strength of the exciton with energy $\Omega_S$ and $A_c$ the area of the supercell.

\subparagraph{Non-defected heterostructure.} The absorbance data for the non-defected heterobilayer was taken from the forthcoming publication \onlinecite{Kleiner2023}, where a GW-BSE calculation in a relaxed supercell of the same size but without vacancy was performed.

\section{Additional results}

\subparagraph{Geometry optimization.}
After relaxation in the presence of the vacancy, we find that close to the defect position, the nearest-neighbor metal-metal distance decreases to $d_{\textnormal{W--W}} = 3.02 $\,\AA\, and $d_{\textnormal{Mo--Mo}} = 3.05 $\,\AA. This corresponds to a reduction of about $3-4$\% in units of the TMDC monolayer lattice constant, $\bar{d}_{\textnormal{X--X}}$. Because the supercell volume is constant during the relaxation, the TMDC layer is strained in the vicinity of the defect where the distance to the next-nearest-neighbor metallic atoms increases up to $\simeq 3.2$ \AA\, for the functional employed here. Similar behavior is also observed for the sulfur atoms surrounding the vacant site, which rigidly follow the motion of the metallic atoms.
Geometry optimization also reduces the average interlayer distance in the heterostructure by around $3-4\%$. 
In particular, we find a reduction of the interlayer distance to $\bar{d}_\textnormal{inter} = 3.29$ \AA\, in the case of MoS\textsubscript{2}--Gr and $\bar{d}_\textnormal{inter} = 3.31$ \AA\, for WS\textsubscript{2}--Gr for our relaxation criteria and our employed van der Waals scheme.

\begin{figure}
   \includegraphics[width=1.0\linewidth]{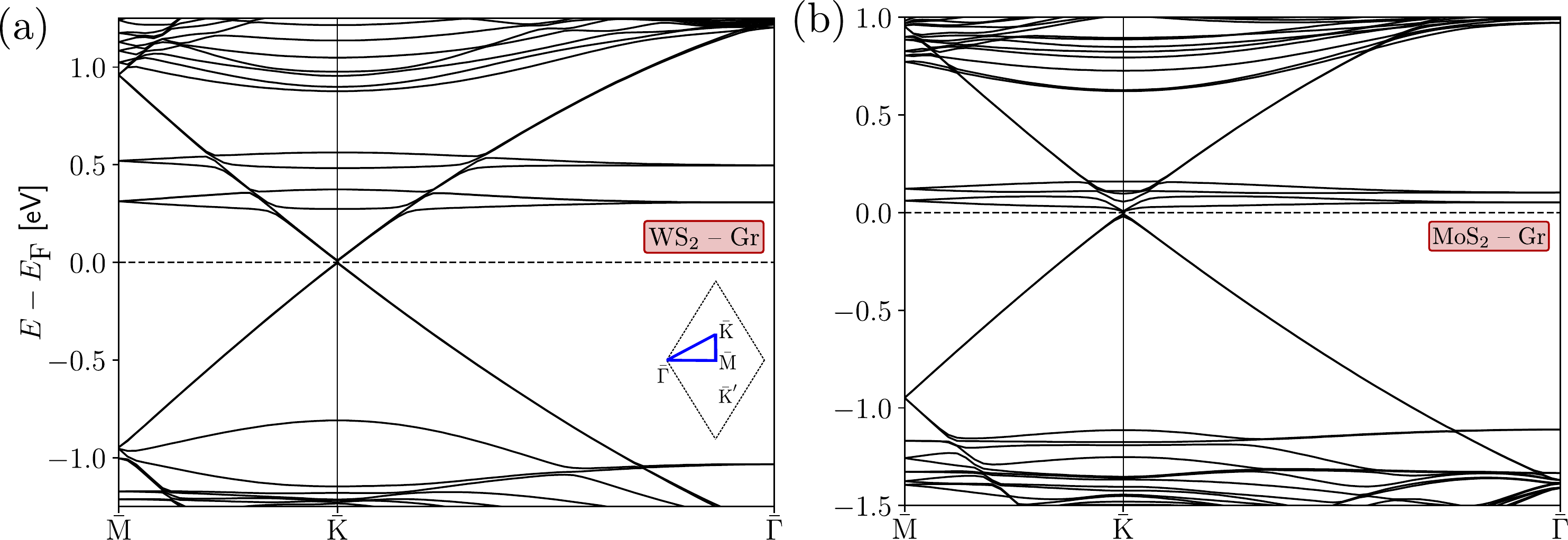}

    \caption{Band structures of the TMDC--Gr heterostructures in the presence of spin-orbit interaction computed along the $\bar{\textnormal{M}}-\bar{\textnormal{K}}-\bar{\Gamma}$ path in the supercell (see inset) (a) WS\textsubscript{2}--Gr heterobilayer (b) MoS\textsubscript{2}--Gr heterobilayer.}\label{f17} 
\end{figure}

\subparagraph{DFT bandstructures.}
For completeness, we show the bandstructures obtained using the DFT relaxed structures in the presence of spin-orbit interaction in Fig. \ref{f17}.

\subparagraph{GW.}
As stated in the main text, our GW results qualitatively follow the DFT band structures discussed in Ref. \onlinecite{Hernangomez2023}. We summarize here for completeness the main features. In essence, the band structures of graphene and the TMDC appear mostly superimposed, with the graphene Dirac cone centered at the $\bar{\textnormal{K}}$ (and $\bar{\textnormal{K}}$') points. The Dirac point sets the charge neutrality point %
within the pristine TMDC band gap. 
The combination of the missing chalcogen atom and the symmetry of the host lattice results in four empty in-gap and two occupied spin-orbit split bands. %
In addition, the reduction of the original TMDC symmetry due to graphene adsorption, and the residual defect-defect interaction that results from the lattice mismatch between layers within the supercell as well as the supercell size (\textit{i.e.} the defect density) make these bands non-degenerate and weakly dispersive. The breaking of the degeneracy is stronger in the vicinity of the $\bar{\textnormal{K}}$ and $\bar{\textnormal{K}}'$ points.
Defect states also hybridize with the graphene states in certain regions of the supercell Brillouin zone. In particular, the interlayer hybridization is relatively important between the defect and the graphene bands in a ring around the $\bar{\textnormal{K}}$ and $\bar{\textnormal{K}}'$ points. This interlayer hybridization was already  measured \cite{Batzill2015, Pierucci2016} and predicted \cite{Jain2018} for occupied bands in MoS\textsubscript{2}--Gr far from the charge neutrality point.

Graphene adsorption and geometry relaxation within the supercell breaks the original lattice symmetry \cite{Hernangomez2023}, lifting the degeneracy of the TMDC conduction states at the $\Lambda$ and K \textbf{k}-points, both folded into the point $\bar{\textnormal{K}}$ of the supercell Brillouin zone due to lattice commensuration \cite{Kleiner2023}. For WS\textsubscript{2}--Gr, this results in the lowest conduction band having $\Lambda$ nature with K states being at higher energy 
already at the DFT level. 
The identification is performed from the pseudo-charge density, $|\Psi|^2$ at $\mathbf{k} = \bar{\textnormal{K}}$, which provides the orbital contribution, comparing  %
to the pristine TMDC states (see also Ref. \onlinecite{Kleiner2023}). 
Screening affects more strongly the more delocalized K states (essentially due to the defect effect), shifting them higher in energy compared to the folded $\Lambda$ states (by $\sim 0.25$ eV), the latter determining thus the pristine band gap. %
For MoS\textsubscript{2}--Gr, %
already at the DFT level the K bands are below the $\Lambda$ bands by $\sim 0.1$ eV and this situation reverses at the GW level, where the K bands appear $\sim 0.15$ eV above. %
This again gives a relative shift of $\sim 0.25$ eV, suggesting that in both heterostructures the levels shifting results from the graphene dielectric screening.

The screening and exchange renormalize also the graphene Fermi velocity associated to the slope of the Dirac cone.
In particular, we obtain an increase of the Fermi velocity of $\sim 34$\% for WS\textsubscript{2}--Gr and $\sim 41$\% for  MoS\textsubscript{2}--Gr, consistent with a reported increase of the Fermi velocity of $\sim 34$\% in Ref. \onlinecite{Li2009} and slightly larger than the $\sim 18\%$ reported in Ref. \onlinecite{Olevano2008} (both calculations for the isolated graphene monolayer).
In addition, we find that for  WS\textsubscript{2}--Gr (resp. MoS\textsubscript{2}--Gr) the Fermi energy shifts from $3.80$ eV in DFT to $4.68$ eV in GW with respect to the vacuum level (resp. $3.60$ eV to $4.50$ eV).
%

\section{Absorption convergence checks}

\begin{figure}
   \includegraphics[width=1.0\linewidth]{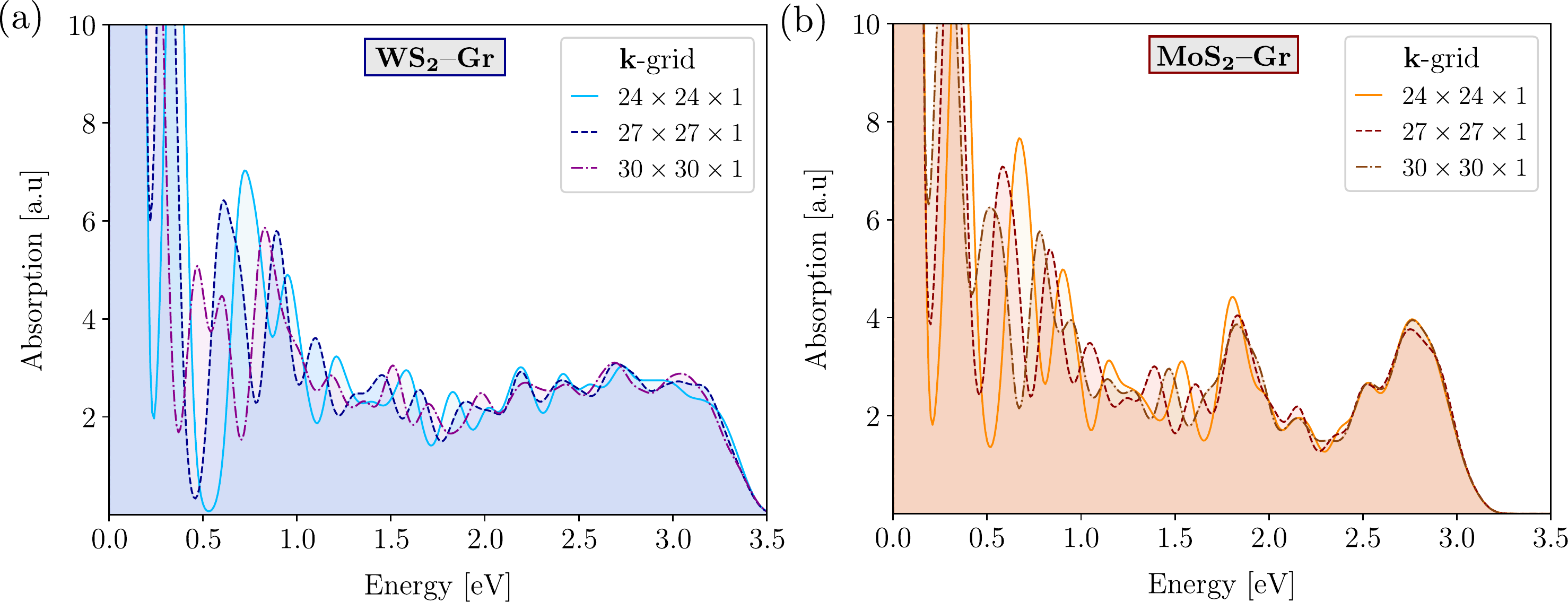}

    \caption{Brillouin zone \textbf{k}-grid convergence of the absorption spectrum for different supercell sampling. (a) WS\textsubscript{2}--Gr heterobilayer (b) MoS\textsubscript{2}--Gr heterobilayer.}\label{f1} 
\end{figure}

\begin{figure}
   \includegraphics[width=0.6\linewidth]{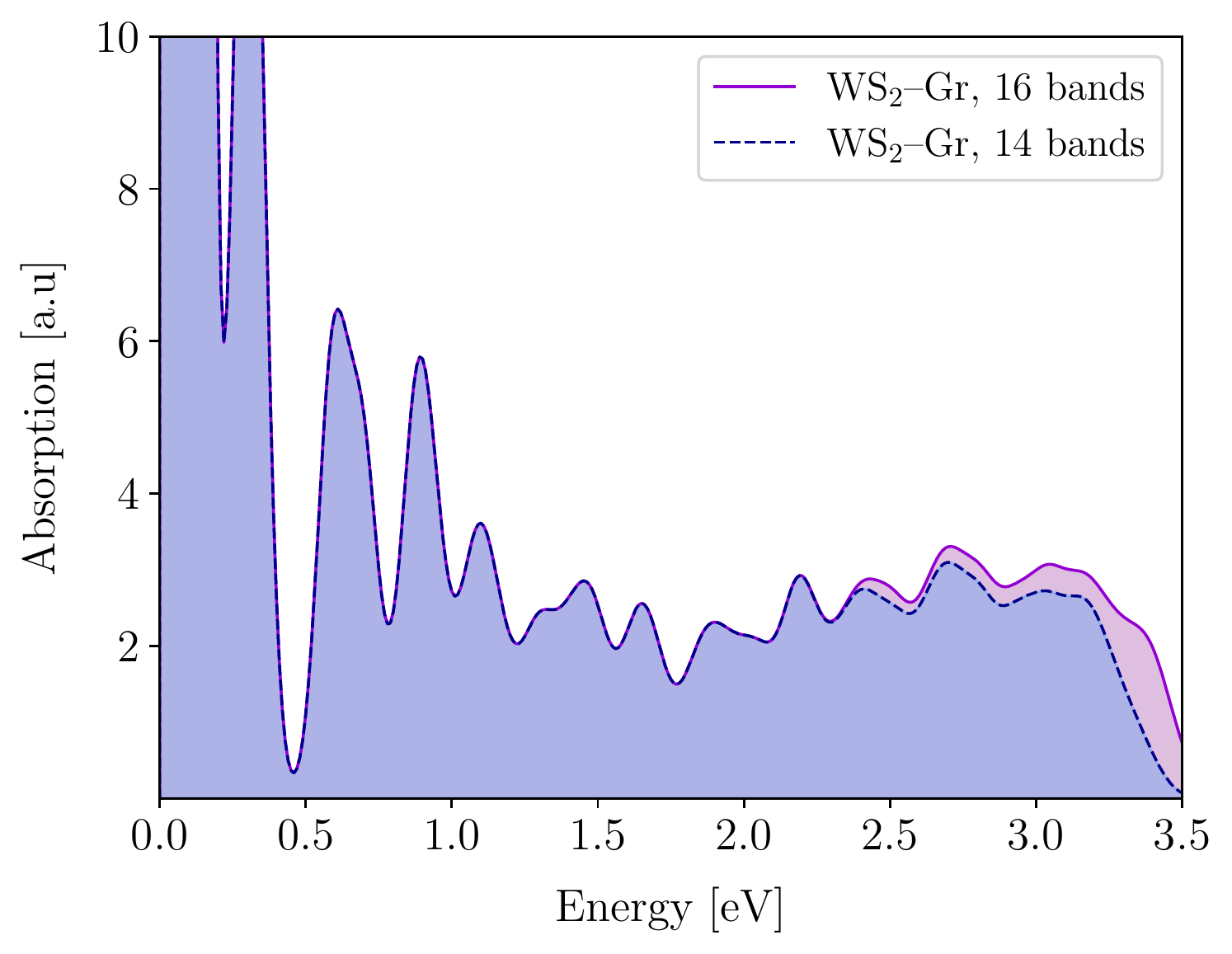}  

    \caption{Convergence of the absorption spectrum of WS\textsubscript{2}--Gr heterobilayer with the number of bands inclued in the BSE calculation.}\label{f16} 
\end{figure}

Ensuring a qualitatively $\mathbf{k}$-grid converged absorption is crucial for understanding the defect-induced phenomena presented in this paper.
In Fig. \ref{f1}, we show the absorption spectrum for different supercell Brillouin zone uniform \textbf{k}-grid samplings of the interpolated absorption grid for (a) WS\textsubscript{2}--Gr (b) MoS\textsubscript{2}--Gr .
We find that the absorption spectra are quantitatively well converged 
in the visible range, especially above optical energies $\gtrsim 2 $ eV for WS\textsubscript{2}--Gr and $\gtrsim 1.75 $ eV for MoS\textsubscript{2}--Gr. In this energy range, convergence is achieved already with a  $24 \times 24 \times 1$ \textbf{k}-grid up to $\sim 10$ meV. However, the position and height of the absorption resonances in the infrared range, especially for energies roughly below $0.3-0.5$ eV, are not yet fully converged. This region, which  is also known to be strongly dominated by intraband graphene resonances \cite{Li2009},  requires very dense \textbf{k}-grids for smoothness and quantitative convergence. 
In the intermediate optical range, up to $\sim 1.6-1.7$ eV, the results are qualitatively converged, but the main absorption features are still quantitatively dependent on the employed \textbf{k}-grid. At   energies between $\sim 0.5-1.6$ eV, the results qualitatively agree between different \textbf{k}-grids, and, importantly, the height of the absorption resonances is stable. The importance of this observation for defect-based sub-gap features and their expected survival in the limit of dense \textbf{k}-sampling is discussed the main text.
Finally, we also ensured that the excitonic features discussed in the main paper are converged in the number of bands included in the BSE calculation, see \textit{e.g.} Fig. \ref{f16}

\section{Absorbance spectra and defect resonances in MoS\textsubscript{2}-graphene}\label{sec:MoS2}

 \begin{figure}
    \includegraphics[width=0.65\linewidth]{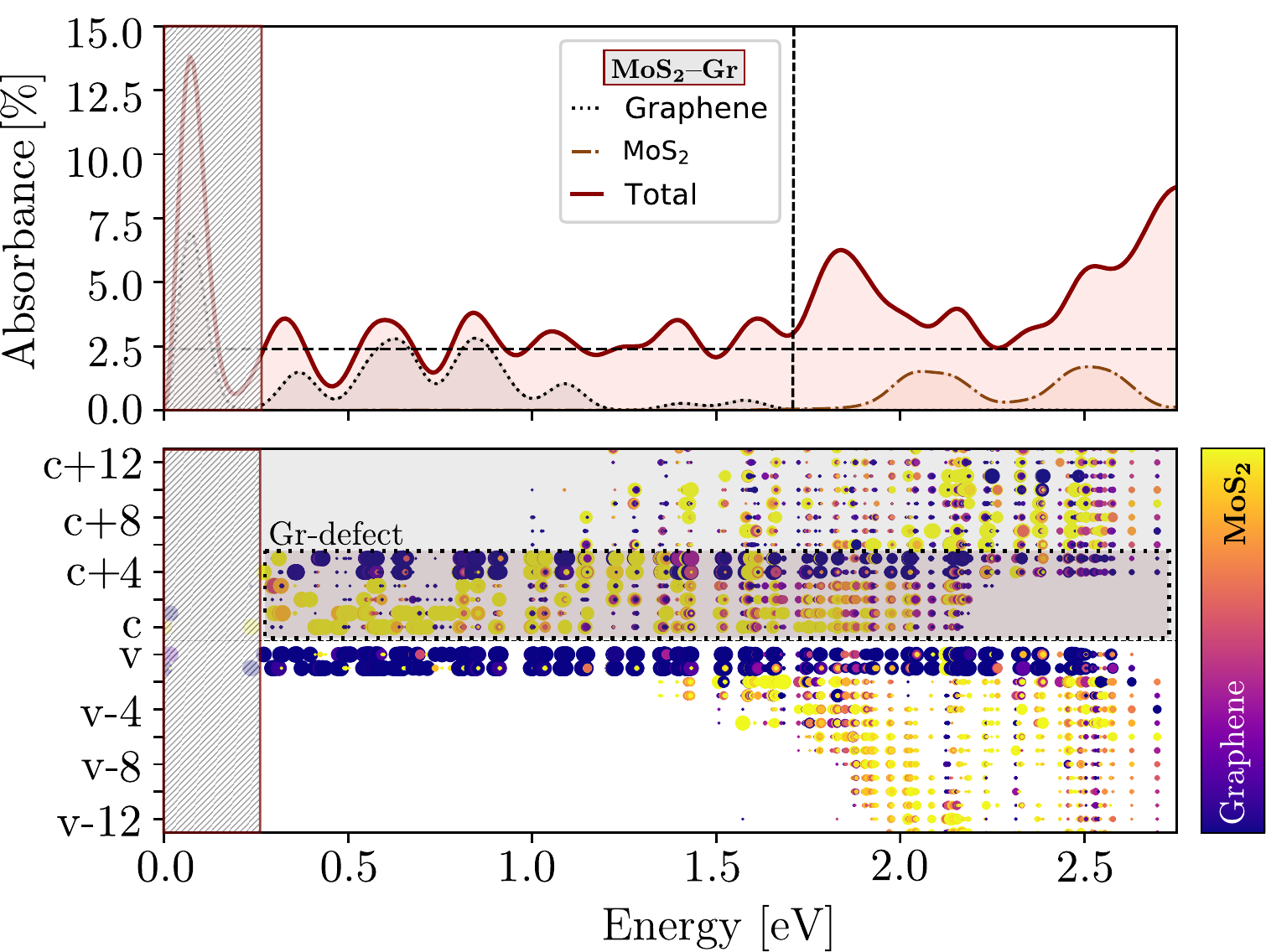} 
     \caption{Absorbance and exciton contributions for the defected MoS\textsubscript{2}--Gr heterobilayer. (Top) Absorbance calculated along one of the main in-plane polarization directions, as well as its decomposition into graphene and MoS\textsubscript{2} contributions (interlayer contributions are read from the difference of the three traces). The dashed horizontal black line marks the $2.4\%$ universal limit of graphene absorbance at infrared energies.
    The shaded box represents the estimated range for which we expect a smooth and monotonic spectrum dominated by graphene (instead of resonances that result from finite \textbf{k}-grid sampling). The vertical dotted line denotes the optical ranges below which excitons transition from being dominated by defect-graphene sub-gap transitions to have transitions which involve pristine TMDC bands.
    (Bottom) For each exciton composing the absorbance resonances, we represent the contribution of each electron and hole bands. Each dot corresponds to the band contribution to a given exciton summed over all $\mathbf{k}$ points (only bright contributions whose oscillator strength are $> 5$ a.u. are shown). For reasons of clarity, all dots with value $\geq 10^3$ a.u. have the same area. 
    The color code corresponds to the layer composition  of each contribution and the dotted box marks the position of the transitions towards graphene-defect empty bands.  
     }\label{f3} 
 \end{figure}
As mentioned in the main text, the strong mixing that occurs in  WS\textsubscript{2}--Gr is also qualitatively similar for the case of the MoS\textsubscript{2}--Gr interface.  We show in Fig. \ref{f3}, top panel, the absorbance spectrum, its decomposition into intralayer graphene, intralayer TMDC and interlayer contributions as well as the graphene universal absorbance limit at infrared energies (dashed horizontal line). Here, however, the impact of the defect bands in the absorbance is more dramatic since the defect bands are closer to the Dirac point and thus interlayer contributions affect the absorbance even at lower energies. Consequently, for MoS\textsubscript{2}--Gr  the absorbance spectrum is very different to that obtained for this TMDC  with chalcogen vacancies in the absence of  graphene layer \cite{Mitterreiter2021, Amit2022}. 
In Fig. \ref{f3}, bottom panel, we show the band contribution of each  exciton. Note that, as compared to WS\textsubscript{2}--Gr, here the four spin-orbit split defect bands can be identified clearly as being below the graphene Dirac cone already at optical energies $\sim 0.5$ eV.

\section{Wavefunction densities of relevant conduction bands}\label{app:wfn}

\begin{figure*}
   \includegraphics[width=1.0\linewidth]{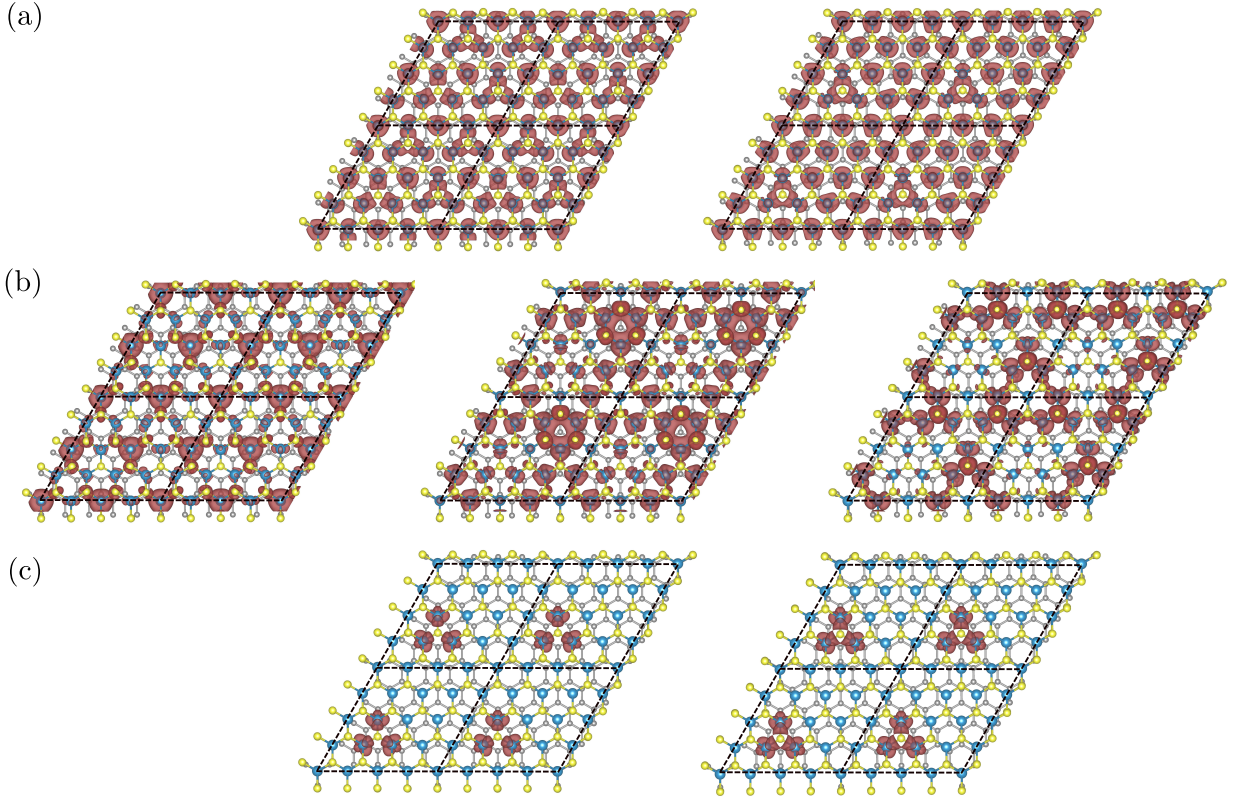}
    \caption{(a) Top view of the Kohn-Sham pseudo-charge density, $|\Psi_{n,\mathbf{k}}(\mathbf{r})|^2$, for the conduction band states, derived from the monolayer K point. These densities are evaluated at the point $\mathbf{k} = \bar{\textnormal{K}}$ of the supercell Brillouin zone of the defected WS\textsubscript{2}--Gr heterobilayer.
    (b) Same as in (a) but for the states derived from the folded $\Lambda$ point, at $\mathbf{k} = \bar{\textnormal{K}}$. (c) Same as in (a) but for the two defect states with different ``orbital'' quantum number. \cite{Hernangomez2023}
    }\label{f4} 
\end{figure*}

In order to understand how the defect has such a strong impact on the absorption features, it is instructive to look at the wavefunction densities of relevant states that participate in the transitions. In Fig. \ref{f4}, we show the pseudo-charge density for three sets of bands at the relevant $\bar{\textnormal{K}}$ point of the supercell Brillouin zone: (a) states with K nature, \textit{i.e.} which have orbital contribution resembling the orbitals found in the monolayer at the K point; (b) states originally coming from the monolayer $\Lambda$ point, folded into $\bar{\textnormal{K}}$, and (c) defect states. 
While the states with K nature are weakly affected by the missing atom forming the vacancy, the perturbation due to the missing atom  affects very strongly the $\Lambda$-like states, which have a new orbital pattern that varies at longer length scales (nanometer scale). As such, we expect these states to react differently to the dielectric screening once this is taken into account. This difference can explain therefore, as stated in the main text, why the pristine TMDC band-gap at the GW level is determined by the states shown in Fig. \ref{f4} (b) and not in (a). For comparison, we display in (c) the pseudo-densities of two defect states. These states present an orbital signature consistent with that seen in panel (b), therefore, we conclude that the behavior and shape of the $\Lambda$-derived states is exclusively determined by the vacancy.

\clearpage

\section{Defect-induced exciton hybridization}

\begin{figure}[h]
   \includegraphics[width=0.575\linewidth]{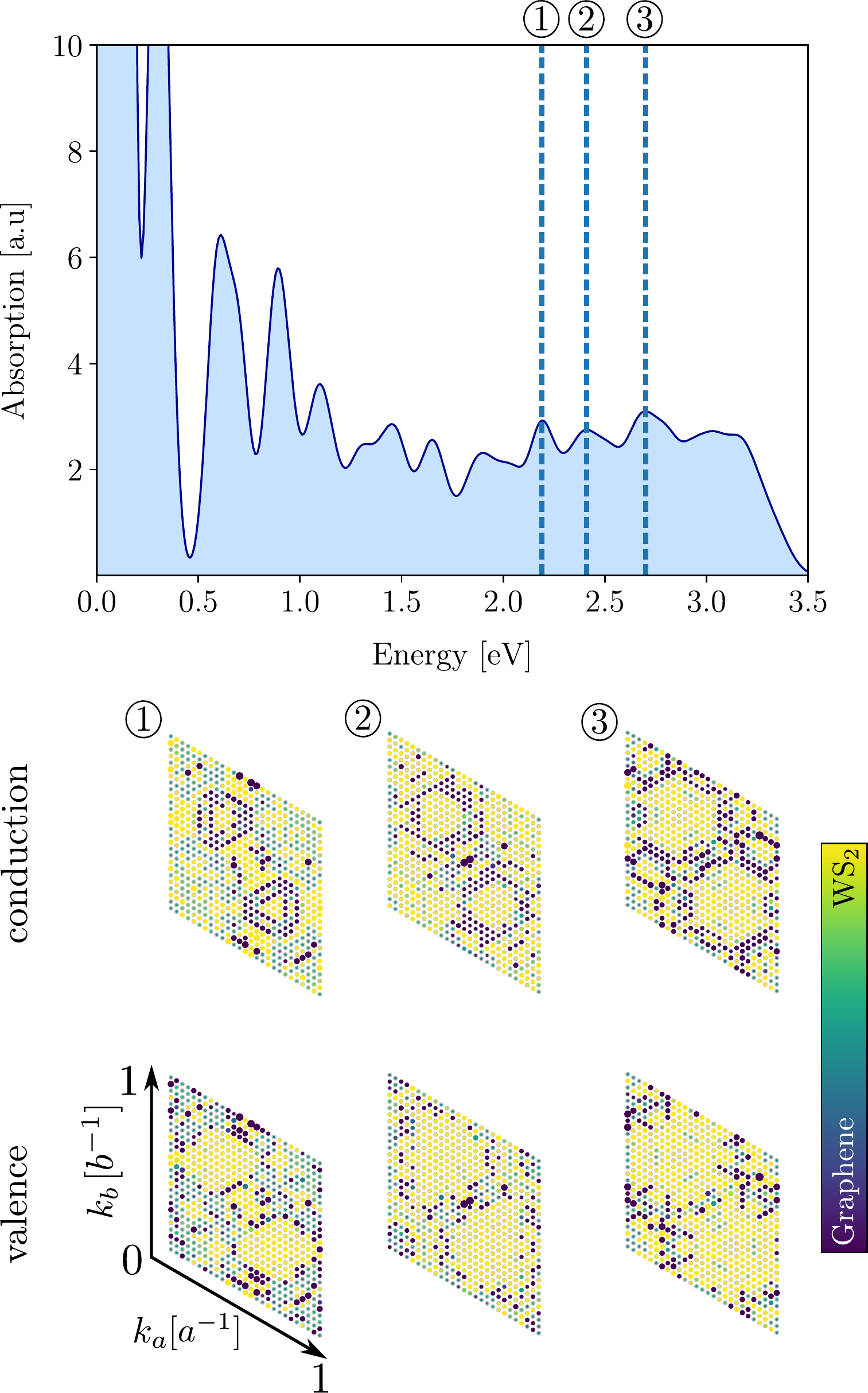} 

    \caption{Brillouin zone exciton distribution plotted for all the WS\textsubscript{2}--Gr excitons within an energy window of $\pm 5 $ meV marked in the absorption spectrum by \textcircled{1} (centered at 2.19 eV), \textcircled{2} (centered at 2.4 eV) and \textcircled{3} (centered at 2.7 eV). Even a small energy window close to the center of the excitation peak shows larger hybridization between the WS\textsubscript{2} and the graphene layers.
    }\label{f18} 
\end{figure}

\begin{figure}[h]
   \includegraphics[width=0.575\linewidth]{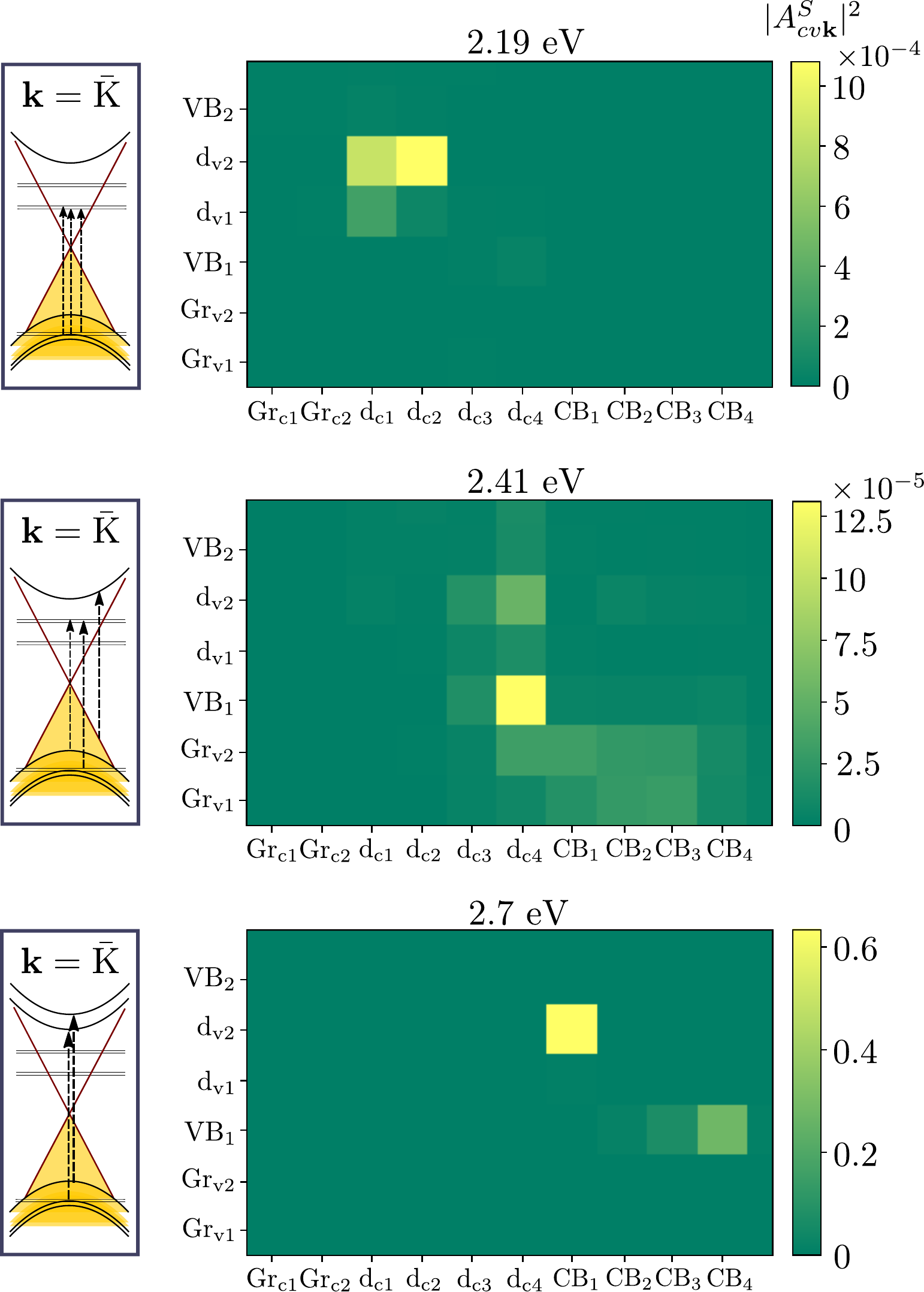} 

    \caption{Sketch of most prominent transitions and transition band diagram for the absorption peaks in Fig. \ref{f18}. The excitons have all been added up within an energy window of $\pm 5 $ meV.
    }\label{f19} 
\end{figure}

\clearpage 

\section{Additional figures}\label{app:add_figs}
\vspace{-0.2cm}
\begin{figure}[h]
   \includegraphics[width=0.5\linewidth]{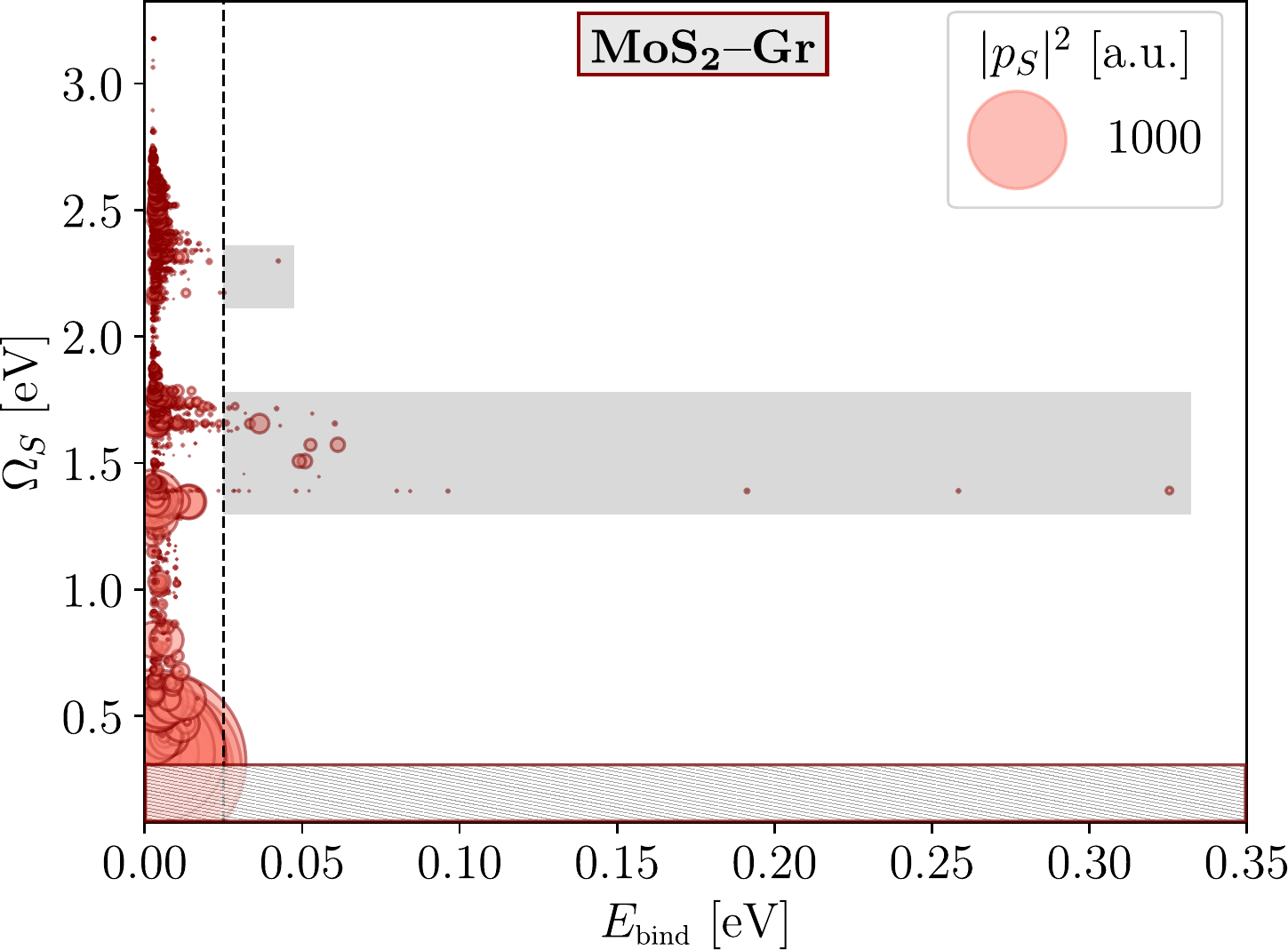}

    \caption{Exciton energies, $\Omega_S$, represented as a function of the binding energy, $E_\textnormal{bind}$, for the excitons in the MoS\textsubscript{2}--Gr heterobilayer.
    The binding energy is computed using Eq. \eqref{eq:binding}. We only show the excitons with $E_\textnormal{bind} > 2.5$ meV (around $\sim 14000$ out of $142 884$ excitons for this \textbf{k}-grid sampling and number of bands). The size of each dot is proportional to the oscillator strength (rescaled by a factor of two for visibility).
    }\label{f6} 
\end{figure}

\begin{figure}[h]
   \includegraphics[width=0.575\linewidth]{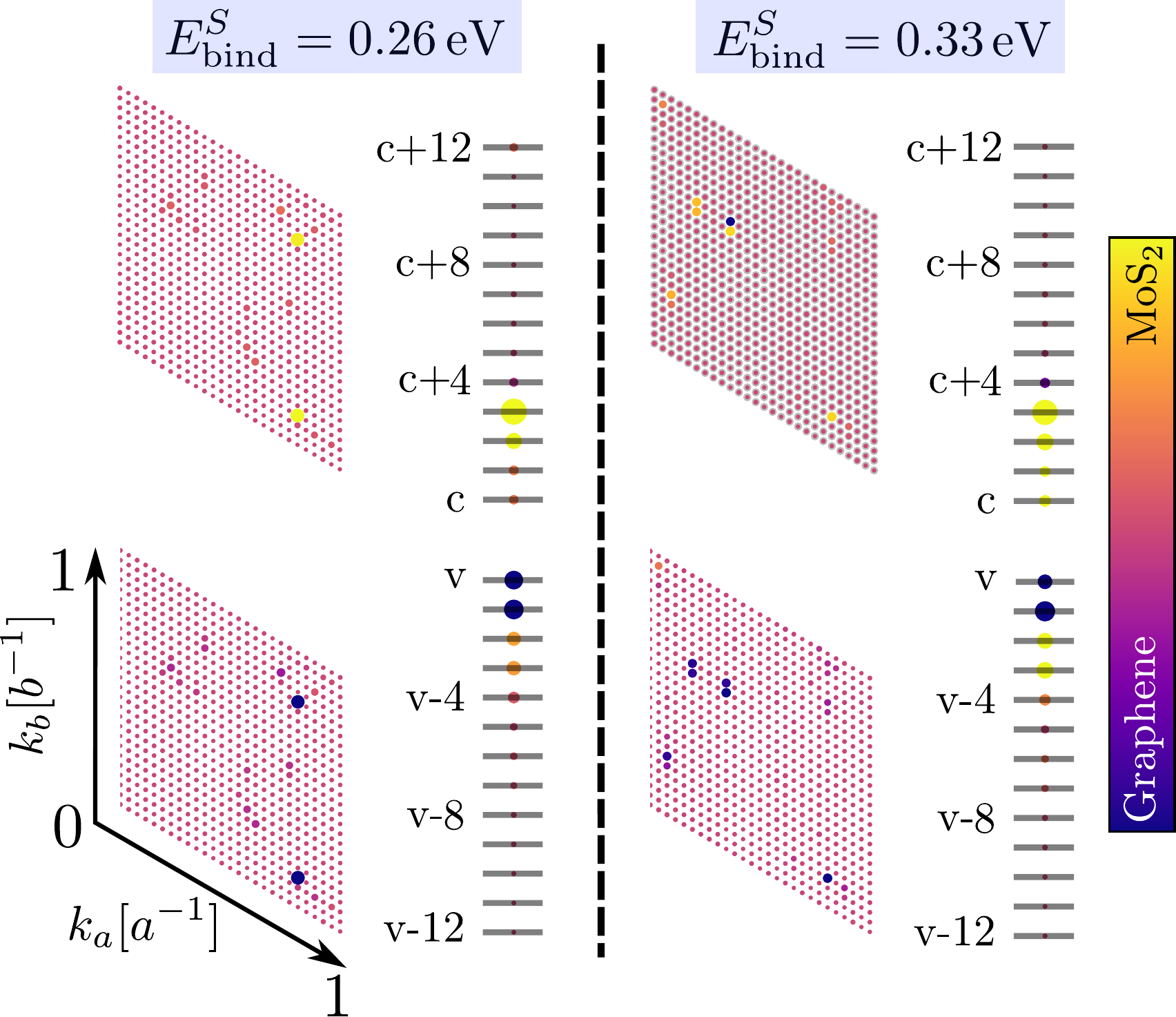} 
    \caption{Brillouin zone exciton distribution for the two most bound excitons  in Fig. \ref{f6}. The Brillouin zone in the top row displays $\sum_{v}|A^S_{cv\mathbf{k}}|^2$, while the lower row shows $\sum_{c}|A^S_{cv\mathbf{k}}|^2$. These two excitons present  graphene-defect transitions and are largely delocalized in \textbf{k}-space.
    }\label{f7} 
\end{figure}

\begin{figure}[h]
   \includegraphics[width=0.6\linewidth]{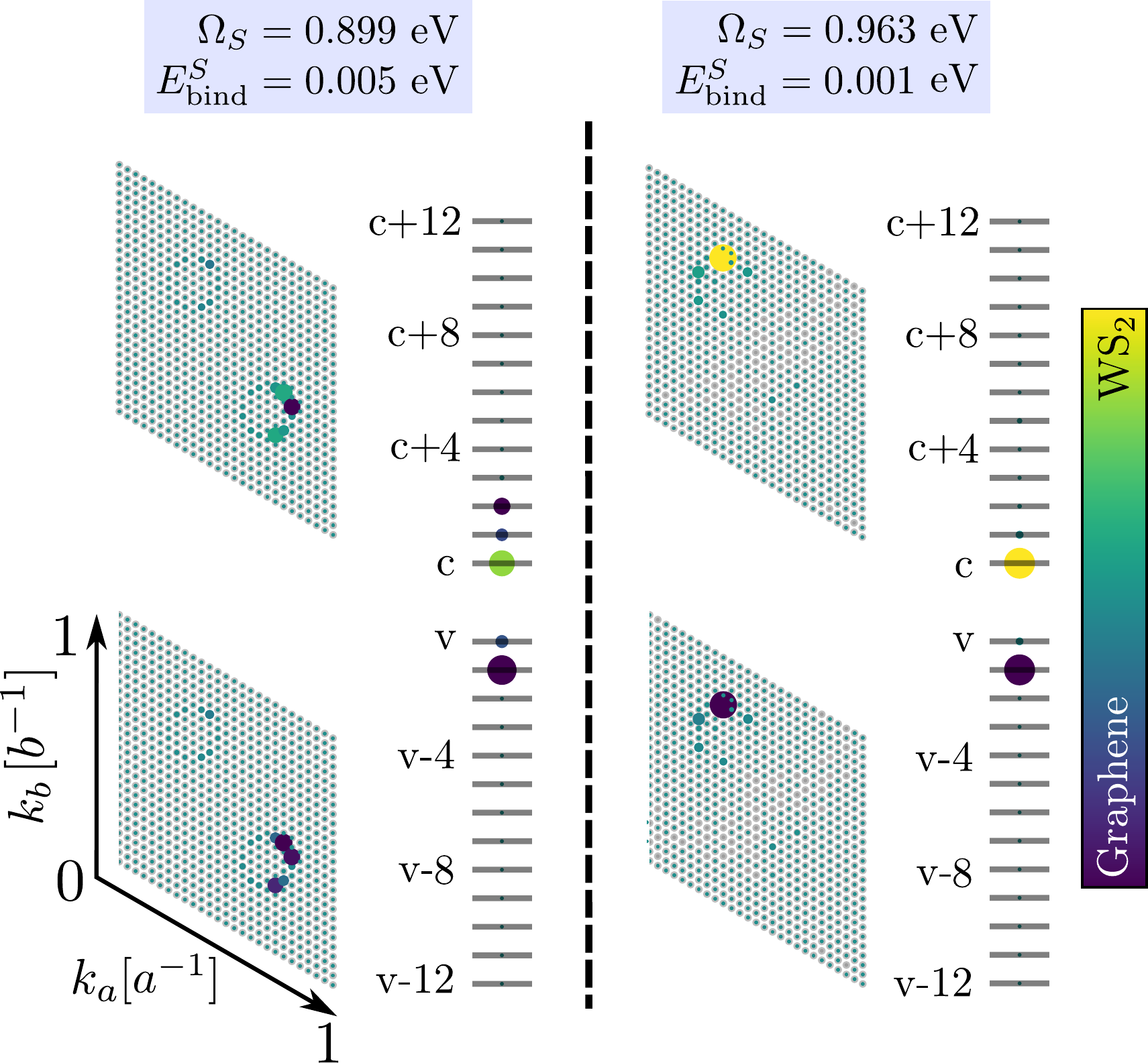} 
    \caption{Same as in Fig. \ref{f7} but for two graphene dominated excitons. Both excitons are mostly localized in the vicinity of the $\bar{\textnormal{K}}$ valleys. These excitons have a large transition dipole, with the value of $\mu_S$ being $1.5 \cdot 10^2$ a.u. and $1.2 \cdot 10^2$ a.u. respectively, but small binding energy.
    }\label{f8} 
\end{figure}

\begin{figure}
   \includegraphics[width=0.6\linewidth]{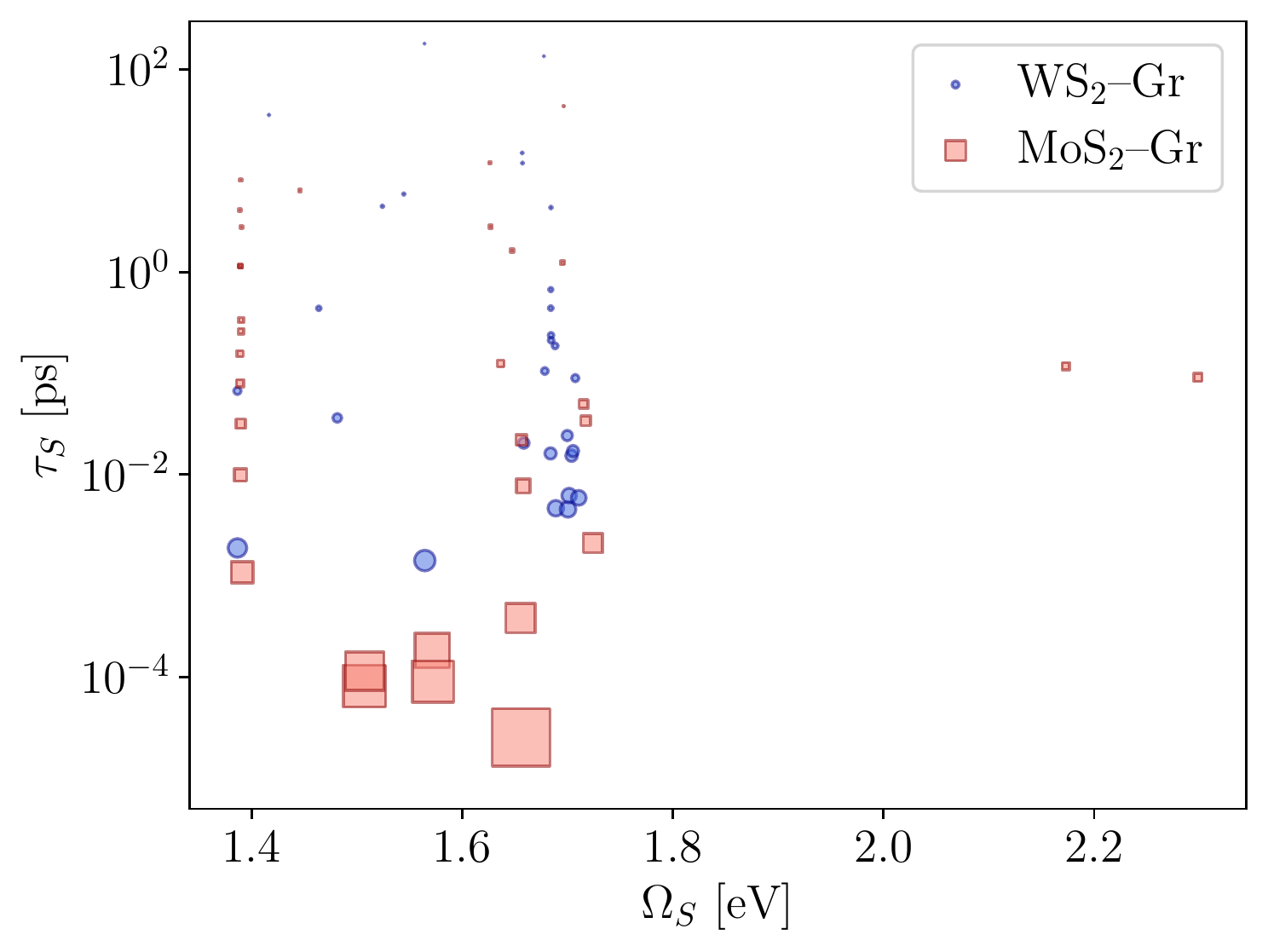}

    \caption{Intrinsic radiative lifetimes at low temperatures for the grey excitons with binding energy larger than $25$ meV, both for WS\textsubscript{2}--Gr (blue circles) and MoS\textsubscript{2}--Gr (red squares). The size of the points is proportional to the oscillator strength, rescaled for clarity by a factor of 20. 
    }\label{f11} 
\end{figure}

\begin{figure}[h]
   \includegraphics[width=0.6\linewidth]{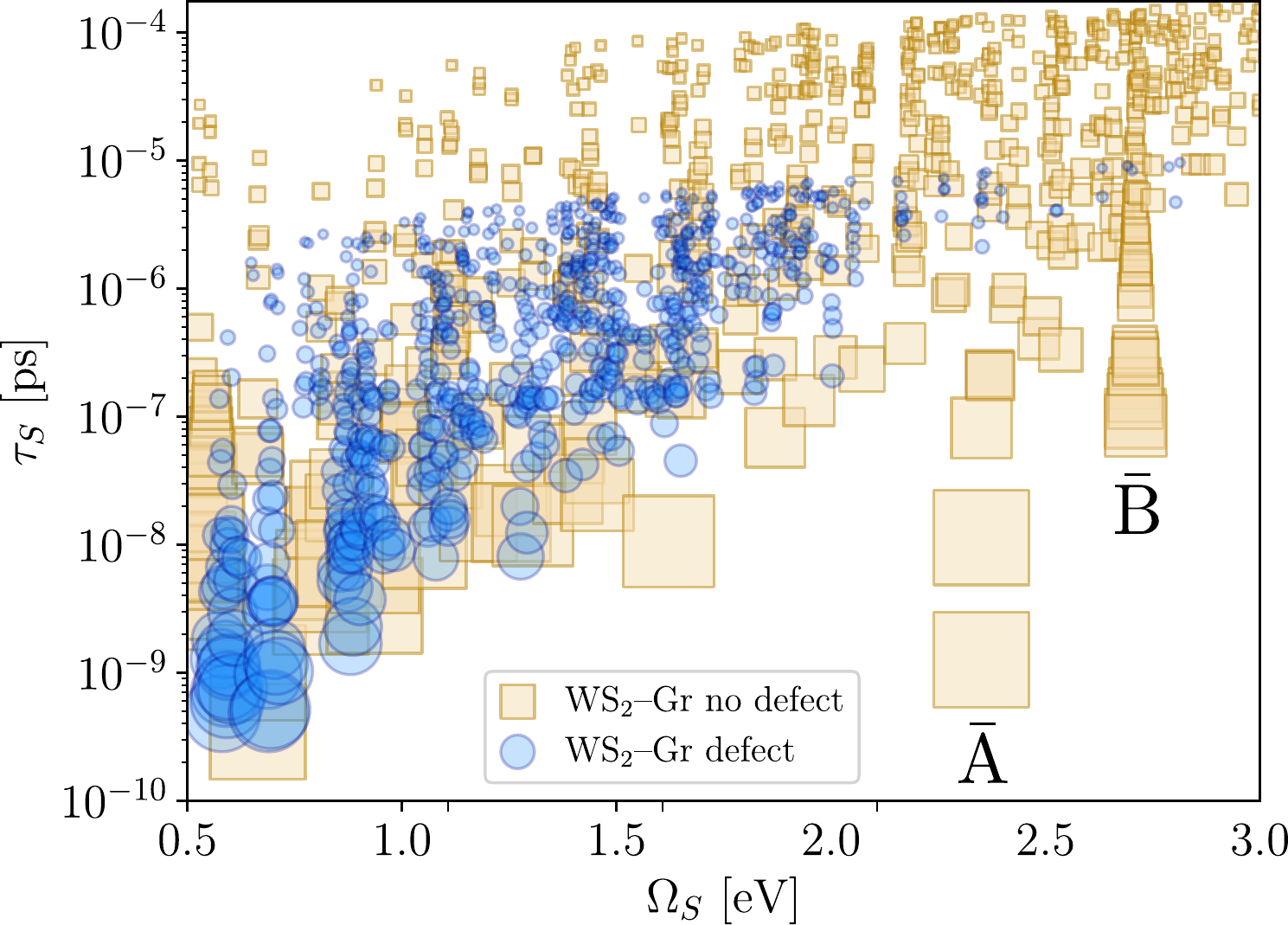} 
    \caption{Intrinsic radiative lifetimes at low temperatures for the bright excitons in the WS\textsubscript{2}--Gr heterobilayer with and without defects (data for the pristine heterobilayer taken from Ref. \onlinecite{Kleiner2023}). The size of the symbols is proportional to the oscillator strength $\mu_S$, which is chosen to be larger than the threshold value of $10^1$ a.u.. For clarity,  we fix the value of $\mu_S$ to be $10^{3}$ a.u if larger or equal. The grey shaded area corresponds to the region of the spectrum strongly dominated by graphene.
    We observe that the heterostructure without the vacancy has brighter excitons with systematically shorter lifetimes and larger oscillator strengths in the visible region, where the defected structure shows exciton quenching. 
    }\label{f9} 
\end{figure}

\clearpage
\bibliography{biblio}